\documentclass[lettersize, final, twoside, journal]{IEEEtran}
\usepackage{cite}
\usepackage{amsmath, amssymb, amsfonts}
\usepackage{enumitem}
\usepackage{algorithmic}
\usepackage{graphicx}
\usepackage{subfigure}
\usepackage{textcomp}
\usepackage{xcolor}
\usepackage{amsthm}
\usepackage{multirow}
\usepackage{bm}
\usepackage{siunitx}
\usepackage[]{hyperref} 
\hypersetup{
hidelinks,
colorlinks=false,
linkcolor=black,
citecolor=black,
}
\allowdisplaybreaks
\usepackage{balance}
\usepackage{longtable, threeparttable}
\usepackage{makecell}

\begin{document}

\title{Generalized Two-Dimensional Index Modulation in the Code–Spatial Domain for LPWAN}

\author{Long Yuan, Wenkun Wen,~\IEEEmembership{Member,~IEEE}, Junlin Liu, Peiran Wu,~\IEEEmembership{Member,~IEEE}, \\ and Minghua Xia,~\IEEEmembership{Senior Member,~IEEE}
	 \thanks{Received 12 January 2026; revised 8 April 2026; accepted 23 April 2026. The associate editor coordinating the review of this article and approving it for publication was H. Lin. \textit{(Corresponding authors: Wenkun Wen; Minghua Xia.)}}
	\thanks{Long Yuan,  Peiran Wu, and Minghua Xia are with the School of Electronics and Information Technology, Sun Yat-sen University, Guangzhou 510006, China (email: yuanlong@mail2.sysu.edu.cn, wupr3@mail.sysu.edu.cn, xiamingh@mail.sysu.edu.cn)}
	\thanks{Wenkun Wen and Junlin Liu are with the R\&D Department, Techphant Technologies Company Ltd., Guangzhou 510310, China (email: wenwenkun@techphant.net, liujunlin@techphant.net)}	
	\thanks{Digital Object Identifier }
}

\markboth{IEEE Transactions on Communications} {Yuan \MakeLowercase{\textit{et al.}}: Generalized Two-Dimensional Index Modulation in the Code–Spatial Domain for LPWAN}

\maketitle

\begin{abstract}
Low-power wide-area networks (LPWANs) are crucial for large-scale Internet of Things (IoT) applications, yet they face increasing demands for higher data rates, improved reliability, and enhanced energy efficiency under stringent hardware constraints. To address these challenges, this paper introduces a generalized code-index modulation (CIM) transceiver that employs multiple-antenna index modulation (IM). The transmitter integrates spatial modulation (SM), space-time block coding (STBC), and CIM into a unified two-dimensional (2D) coding structure, where the spreading sequences--realized via continuous phase modulation with spread spectrum (CPM-SS), chirp spread spectrum, or Zadoff--Chu sequences---serve as spreading codes. Three specific schemes are proposed: SM-CIM, STBC-SM-CIM, and an enhanced STBC-SM-CIM (ESTBC-SM-CIM), designed to jointly improve data rate and energy efficiency. Closed-form expressions for the average bit error probability are derived, and system performance is analyzed in terms of data rate, energy efficiency, and computational complexity. Simulation results show that the proposed designs consistently outperform benchmark schemes, demonstrating their potential for enabling high-data-rate, energy-efficient LPWAN and IoT communications.
\end{abstract}
 

 \IEEEpubidadjcol

\section{Introduction}
\IEEEPARstart{T}{he} Internet of Things (IoT) is transforming everyday life by enabling greater convenience, operational efficiency, and immersive intelligence. However, the large-scale deployment of IoT networks introduces significant challenges, including the need to support massive device connectivity, minimize deployment and operational costs, and ensure ultra-low power consumption. Moreover, IoT applications often require long-range, reliable communication and extensive coverage—requirements that conventional wireless technologies, such as wireless local area networks (WLANs) and cellular networks, struggle to meet due to high infrastructure costs and limited coverage areas \cite{LPWA}. To overcome these limitations, low-power wide-area network (LPWAN) technologies have emerged as a promising solution, offering cost-effective, energy-efficient connectivity tailored for large-scale IoT scenarios \cite{LPWAN-LoRa-NBIoT}. These advantages make LPWAN particularly well-suited for application domains such as smart cities, logistics, and agriculture, where low power and low cost are critical \cite{LPWAN-2023}.

Over the past decade, several representative LPWAN solutions have been developed. Narrowband IoT (NB-IoT) employs orthogonal frequency-division multiplexing (OFDM) in licensed spectrum \cite{NBIoT}; LoRa uses frequency shift chirp spread spectrum (FSCSS) in unlicensed bands \cite{LoRa}; and Sigfox adopts ultra-narrowband differential binary phase-shift keying (D-BPSK) \cite{Sigfox}. Additionally, the IEEE 802.15.4 standard extends support for low-power wireless personal area networks in IoT \cite{2024-IEEE-standard}. Despite their broad adoption, these technologies still face limitations in terms of data rate, interference resilience, and energy efficiency. As IoT continues to evolve, there is an urgent need for advanced physical-layer solutions that preserve LPWAN's low-complexity characteristics while delivering improved performance.

Index modulation (IM) has gained increasing attention as a physical-layer technique that can improve both energy efficiency and data rates—key requirements for next-generation wireless communications \cite{IM-NextG}. IM partitions the transmitted bitstream into two components: one part determines which communication resource (e.g., antenna, subcarrier, time slot, or spreading code) is activated, while the other part is modulated using conventional schemes such as phase shift keying (PSK) or quadrature amplitude modulation (QAM) \cite{Multidim-IM}. The resource index bits are implicitly transmitted through the selection process, incurring no additional energy expenditure. This dual functionality makes IM inherently power-efficient and spectrally efficient. For instance, spatial modulation (SM) outperforms conventional multiple-input multiple-output (MIMO) systems in energy-constrained scenarios \cite{SM}; orthogonal frequency division multiplexing with index modulation (OFDM-IM) exhibits superior performance over conventional OFDM \cite{OFDM-IM}; and code index modulation (CIM) has been shown to surpass traditional direct-sequence spread spectrum (DSSS) techniques \cite{CIM}. These advantages position IM as a strong candidate for enhancing the performance of LPWAN.

 \IEEEpubidadjcol
 
Recent research has explored the fusion of IM with LPWAN-friendly techniques, particularly within the LoRa and chirp spread spectrum (CSS) paradigms. Proposals include FSCSS-IM \cite{FSCSS-IM}, in-phase/quadrature chirp index modulation (IQ-CIM) \cite{IQ-CIM}, spreading factor (SF) based IM for LoRa \cite{IM-LoRa-SF}, and group-based CSS modulation techniques \cite{IQ-GCSS}. Furthermore, Zadoff-Chu (ZC)-based Golden modulation \cite{Golden-Modulation-ZC-IoT} and continuous phase modulation with spread spectrum (CPM-SS) \cite{Iot-zhangruiqi, CPM-SS-Long, wen2025unifiedblocksignalprocessing} have been investigated to improve robustness and spectral efficiency. Importantly, all these orthogonal or quasi-orthogonal sequences (e.g., FSCSS, ZC, and CPM-SS) can serve as candidate spreading codes for CIM to meet the low-power and low-cost requirements of LPWAN. However, most of these schemes are constrained to a single-antenna configuration, which limits the achievable diversity and data-rate gains. 

Meanwhile, MIMO technologies have been widely adopted to enhance link reliability, spectral efficiency, and data rates in cellular communications, and their application to IoT scenarios is growing \cite{5G-IoT}. Prior work has explored MIMO techniques in LPWAN systems, including STBC-MIMO LoRa \cite{STBC-MIMO-LoRa}, data-rate-enhanced MIMO-LoRa \cite{MIMO-LoRa-data-rate}, and 1-Bit-STBC-LoRa for extreme conditions \cite{1-Bit-STBC-LoRa}. However, conventional MIMO architectures entail high power consumption and computational complexity, thereby limiting their applicability in low-power IoT environments. To bridge this gap, SM offers a simplified MIMO alternative that achieves energy savings by activating only one antenna per time slot. When combined with space-time block coding (STBC), the resulting STBC-SM framework can achieve both diversity and energy efficiency \cite{STBC-SM}. Enhancements like quasi-orthogonal STBC-SM \cite{QOSTBC-SM} and temporal-permutation STBC-SM \cite{TP-STBC-SM} have been proposed. However, most of these solutions are limited to spatial-domain exploitation and fail to leverage other dimensions, such as code-domain diversity. 

To further enhance IM performance, recent work has proposed two-dimensional (2D) schemes, including the generalized CIM-SM (GCIM-SM) \cite{GCIM-SM} and the generalized space-code index modulation (GSCIM) \cite{GSCIM}. Nevertheless, these schemes did not fully exploit transmit diversity, which is an effective method for enhancing the reliability of signal transmission in fading environments. Moreover, relatively few studies combine physical-layer modulation techniques (e.g., FSCSS, ZC, and CPM-SS) of LPWAN with SM and STBC to enhance system performance.

To address the aforementioned limitations, this paper proposes a generalized multiple-antenna IM transceiver that integrates CIM, SM, and STBC. Specifically, we present three schemes: \textbf{SM-CIM}, \textbf{STBC-SM-CIM}, and \textbf{enhanced STBC-SM-CIM (ESTBC-SM-CIM)}. These designs combine the property of the spreading sequences with the energy efficiency of SM and the diversity gain of STBC, providing an LPWAN-oriented solution that reduces power consumption and costs while improving reliability and data rates. The key contributions of this work are summarized as follows:
\begin{itemize}
	\item A generalized transceiver is designed, comprising three practical schemes: SM-CIM, STBC-SM-CIM, and ESTBC-SM-CIM. The transmitter employs spreading sequences (e.g., CPM-SS, FSCSS, and ZC) and antenna selection to construct a 2D space-time index modulation structure tailored to LPWAN requirements. 
	\item A low-complexity (LC) despreading-based detection algorithm is developed at the receiver, which reduces computational overhead while ensuring reliable demodulation in resource-constrained IoT devices. 
	\item The average bit error probability (ABEP) of the proposed schemes is analytically derived and evaluated. In addition, data rate, energy efficiency, and computational complexity are analyzed and compared with existing approaches. Extensive simulation and analytical results demonstrate the superiority of the proposed schemes over benchmark methods in terms of bit error rate (BER), data rate, and energy efficiency, thereby highlighting their potential for practical LPWAN applications that require enhanced system efficiency and reliability.
\end{itemize}

The remainder of this paper is organized as follows: Section~\ref{Section-II} presents the transceiver designs for the proposed SM-CIM, STBC-SM-CIM, and ESTBC-SM-CIM schemes. In Section~\ref{Section-III}, the performance analysis of the proposed schemes is conducted. Section~\ref{Section-IV} presents and discusses the simulation results. Finally, Section~\ref{Section-V} concludes the paper.

{\it Notation:}  The superscripts $(\cdot)^T$, $(\cdot)^H$, and $(\cdot)^{\ast}$ denote the transpose, conjugate transpose, and complex conjugate operations, respectively. The symbol $\bm{0}$ denotes a vector of all entries being zero. The operators $\operatorname{vec}(\cdot)$, $\mathbb{E}[\cdot]$, ${\Re}(\cdot)$, $(\cdot)!$, $\lfloor \cdot \rfloor$, and $\lceil \cdot \rceil$ denote vectorization, expectation, real part, factorial, floor, and ceiling operations, respectively. The operators $|\cdot|$, $\Vert \cdot \Vert_2$, $\operatorname{Tr}(\cdot)$, $\Vert \cdot \Vert_F$, and $\det(\cdot)$ denote the magnitude, the Euclidean $\ell_2$-norm, trace of a matrix, the Frobenius norm, and determinant of a matrix, respectively. The imaginary unit is denoted by $j \triangleq \sqrt{-1}$. The notation $\bigcup$, $\sum$, and $\prod$ denote the union, summation, and product, respectively. The notation $\mathcal{CN}(\mu, \sigma^2)$ denotes a complex Gaussian distribution with mean $\mu$ and variance $\sigma^2$. The binomial coefficient is defined as $\binom{n}{k} \triangleq \tfrac{n!}{k!(n-k)!}$. The notation $\Pr(\cdot)$ denotes the probability of an event. The special function $M_u(x) \triangleq \int_{0}^{\infty} f_u(u) e^{x u} \mathrm{d}u$ denotes the moment generating function (MGF) of the random variable $u$, where $f_u(u)$ is the probability density function (PDF) of $u$. The Gaussian $Q$-function is defined as $Q(x) \triangleq \tfrac{1}{\sqrt{2\pi}} \int_{x}^{\infty} \exp\left(-{u^2}/{2 }\right) \mathrm{d}u$. 

\section{Transceiver Design} \label{Section-II}
This section first presents the basic system models of SM, STBC-SM, and CIM. Next, the transceiver designs of the proposed SM-CIM, STBC-SM-CIM, and ESTBC-SM-CIM schemes are introduced. We consider a widely adopted MIMO channel model in which the transmitter and receiver are equipped with $N_t$ transmit and $N_r$ receive antennas, respectively. The channel is assumed to follow independent and identically distributed (i.i.d.) frequency-flat Rayleigh fading. The channel matrix is denoted by $\bm{H} \in \mathbb{C}^{N_r \times N_t}$, whose entries are modeled as circularly symmetric complex Gaussian random variables with zero mean and unit variance.

\subsection{The Basic System Model}
\subsubsection{SM} 
In SM, the information bits are partitioned into two parts. The first part comprises index bits that select the active transmit antenna, while the second part is mapped to a conventional modulation symbol, such as PSK or QAM \cite{Survey-SM}. Accordingly, the total number of bits conveyed per symbol interval is given by $b_t = \log_2 N_t + \log_2 M$, where $M$ denotes the constellation size. The transmit signal vector of SM can be expressed as
\begin{equation} \label{eq-SM-signal}
   \bm{x}_{\text{SM}} = [0, \cdots, 0, x, 0, \cdots, 0]^T,
\end{equation}
where $x$ is the $M$-ary modulated symbol, and the only non-zero element is located at the $n_t^{\text{th}}$ position corresponding to the selected active antenna, with $n_t \in \{1, 2, \cdots, N_t\}$.

\subsubsection{STBC-SM}
The STBC-SM scheme utilizes Alamouti's STBC, where two complex constellation symbols, $x_1$ and $x_2$, are transmitted orthogonally from two selected transmit antennas over two consecutive symbol intervals \cite{STBC-SM}. The corresponding codeword is expressed as 
\begin{equation} \label{eq-STBC-SM-codeword}
   \bm{X}_0 = \begin{pmatrix}
      \bm{x}_1^T \\
      \bm{x}_2^T
      \end{pmatrix} = 
      \begin{pmatrix}
         x_1 & x_2 \\
         -x_2^{\ast} & x_1^{\ast} \\
      \end{pmatrix}, 
\end{equation}
where the rows and columns represent the symbol intervals and transmit antennas, respectively. 

Following the design methodology in \cite{STBC-SM}, the total number of STBC-SM codewords is given by $N_z = \left\lfloor \log_2{\binom{N_t}{2}} \right\rfloor$. Each codebook $\mathcal{X}_q \in \mathbb{C}^{N_t \times 2}$ contains $N_v = \left\lfloor N_t / 2 \right\rfloor$ codewords, where $q = 1, 2, \cdots, N_b$. The total number of codebooks is $N_b = \left\lceil N_z / N_v \right\rceil$, and the complete codeword set is represented as $\mathcal{X} = \bigcup_{q=1}^{N_b} \mathcal{X}_q$. To maximize the coding gain distance (CGD), the rotation angle $\varphi_q$ for each codebook $\mathcal{X}_q$ is optimized by solving:
\begin{subequations}
\begin{align}
   \delta_{\min}(\bm{X},\bm{\hat{X}}) &= \min \ \det \left((\bm{X} - \bm{\hat{X}}) (\bm{X} - \bm{\hat{X}})^H\right),  \\
   \delta_{\min}(\mathcal{X}_q,\mathcal{X}_p) &= \min_{k,l} \ \delta_{\min}(\bm{X}_{q,k},\bm{X}_{p,l}), \\
   \delta_{\min}(\mathcal{X}) &= \min_{q \neq p} \ \delta_{\min}(\mathcal{X}_q,\mathcal{X}_p), \\
   \bm{\varphi}_{\text{op}} &= \arg\max_{\bm{\varphi}} \ \delta_{\min}(\mathcal{X}), 
\end{align}
\end{subequations}
where $\bm{X}$ and $\bm{\hat{X}}$ denote any two STBC-SM codewords, $\bm{X}_{q,k}$ represents the $k^{\text{th}}$ codeword in the $q^{\text{th}}$ codebook, $\bm{X}_{p,l}$ represents the $l^{\text{th}}$ codeword in the $p^{\text{th}}$ codebook, and $\bm{\varphi} = [1, \varphi_2, \varphi_3, \cdots, \varphi_{N_b}]^T$ denotes the vector of rotation angles for all codebooks. Depending on the number of transmit antennas, the rotation angles can be determined as follows \cite{STBC-SM}: 
\begin{itemize}
   \item {\it Case 1} ($N_t \leq 4$): There are only two codebooks with one non-zero rotation angle. For PSK constellations, the optimal rotation angles obtained via computer search are $\varphi = 1.57$, $0.61$, $0.30$, and $0.15$ \si{\radian} for modulation orders $M = 2, 4, 8$, and $16$, respectively.
   \item {\it Case 2} ($N_t > 4$): For larger antenna arrays, the optimal rotation angles are calculated as 
   \begin{equation} \label{eq-STBC-SM-rotation-angle}
         \varphi_k = \frac{(k-1)\pi}{N_b}, \frac{(k-1)\pi}{2N_b}, \frac{(k-1)\pi}{4N_b}, \frac{(k-1)\pi}{8N_b}, 
   \end{equation}
   where $k = 1, 2, \cdots, N_b$ is the rotation angle index.
\end{itemize}

\begin{table}[!t]
    \renewcommand\arraystretch{1.3}
    \centering
    \caption{Basic Parameters and Antenna Pair of STBC-SM}
    \label{tab-parameter-STBC-SM}
    \begin{tabular}{|c|c|c|c|>{\centering\arraybackslash}m{0.38\linewidth}|}
       \hline
        $N_t$ & $N_z$ & $N_v$ & $N_b$ & Antenna Pair \\
       \hline
        3 & 2 & 1 & 2 & (1,2), (2,3) \\
       \hline
        4 & 4 & 2 & 2 & (1,2), (3,4), (2,3), (4,1) \\
       \hline
        5 & 8 & 2 & 4 & (1,2), (3,4), (2,3), (4,5), (1,3), (2,4), (3,5), (4,1) \\
       \hline
        6 & 8 & 3 & 3 & (1,2), (3,4), (5,6), (2,3), (4,5), (6,1), (1,3), (2,4) \\
       \hline
        7 & 16 & 3 & 6 & (1,2), (3,4), (5,6), (2,3), (4,5), (6,7), (1,3), (2,4), (5,7), (1,4), (2,5), (3,6), (1,5), (2,6), (3,7), (1,6) \\
       \hline
        8 & 16 & 4 & 4 & (1,2), (3,4), (5,6), (7,8), (2,3), (4,5), (6,7), (8,1), (1,3), (2,4), (5,7), (6,8), (1,5), (2,6), (3,7), (4,8) \\
       \hline
    \end{tabular}
\end{table}

The following design criteria govern the antenna-pair selection for index mapping:
\begin{enumerate}
\item For each codeword in the codebook, a distinct transmit-antenna pair is assigned;
\item Antenna pairs are not reused across different codebooks;
\item The degree of antenna overlap between any two codewords is minimized.
\end{enumerate}
These criteria are imposed to reduce inter-codeword interference during detection, thereby improving overall error performance.

Table~\ref{tab-parameter-STBC-SM} summarizes the basic system parameters and the corresponding antenna-pair assignments for different numbers of transmit antennas. The antenna-pair combinations listed in Table~\ref{tab-parameter-STBC-SM} are mapped to the codewords $\bm{X}_{1}$ through $\bm{X}_{N_z}$ in a left-to-right and top-to-bottom order.

\subsubsection{CIM}
CIM partitions the transmitted bits into two components: modulation bits and code-index bits. The code-index bits are used to select a spreading sequence from a predefined codebook, while the modulation bits are mapped onto a conventional constellation such as PSK or QAM. Accordingly, the total number of bits conveyed per CIM symbol is given by $b_t = \log_2{N_c} + \log_2{M}$, where $N_c$ denotes the number of available spreading sequences and $M$ is the constellation size. It is worth noting that, in contrast to \cite{CIM}, the proposed scheme employs the same spreading sequence for both the in-phase and quadrature components of each constellation symbol. The signal expression sent by CIM is as follows:
\begin{equation} \label{eq-CIM-signal}
   \bm{x}_{\text{CIM}} = \bm{z}_{n_c} x,
\end{equation}
where $\bm{z}_{n_c}$ is the selected spreading sequence and $x$ is the modulated symbol.

In this paper, we generalize the notion of spreading codes to encompass a broader class of spreading sequences, treating FSCSS, ZC, and CPM-SS sequences within a unified code-domain framework.

\subsection{SM-CIM Scheme} \label{Section-II-A}
\begin{figure*}[!t]
\centerline{\includegraphics[width=1.0\textwidth, height=0.2\textwidth]{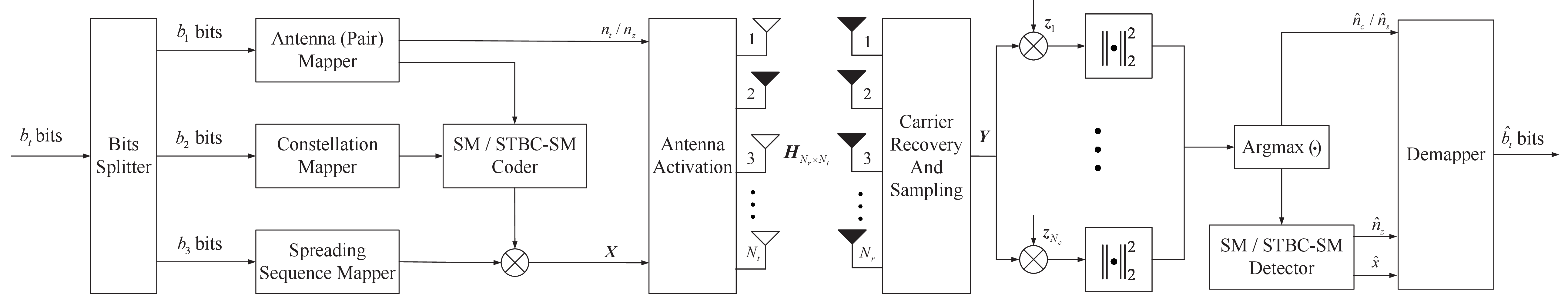}}
\caption{Block diagram of the generalized LC transceiver including three themes: SM-CIM, STBC-SM-CIM, and ESTBC-SM-CIM.}
\label{fig-GMAIM-STBC-SM-transceiver}
\end{figure*}

\subsubsection{Transmitter}
As shown in Fig.~\ref{fig-GMAIM-STBC-SM-transceiver}, the transmitter of the proposed SM-CIM scheme consists of the following processing stages. First, the input bit stream of length $b_t$ is partitioned into three groups: $b_1$ antenna index bits, $b_2$ constellation modulation bits, and $b_3$ code index bits. The $b_1$ bits are converted to decimal to determine the index of the active transmit antenna. The $b_2$ bits are mapped onto an $M$-ary PSK constellation symbol $x$. The $b_3$ bits are converted into a decimal value to select a spreading sequence from the predefined codebook $\bm{Z} \in \mathbb{C}^{L \times N_c}$, where $N_c$ is the total number of spreading sequences and $L$ denotes the sequence length. The selected spreading sequence is denoted as $\bm{z}_{n_c} \in \mathbb{C}^{L \times 1}$, where $n_c \in \{1, 2, \cdots, N_c\}$ represents the sequence index. Next, the modulated symbol $x$ is processed by the SM encoder to generate the spatially modulated signal. Finally, the selected spreading sequence $\bm{z}_{n_c}$ is applied to the SM signal to obtain the transmitted signal $\bm{X}$, which can be expressed as 
\begin{equation} \label{eq-SM-CIM-transmitted-signal}
   \bm{X} = [\bm{0}, \cdots, \bm{0}, \underset{\uparrow \ n_t}{\bm{z}_{n_c} x}, \bm{0}, \cdots, \bm{0}]^T,
\end{equation}
where $\bm{X} \in \mathbb{C}^{N_t \times L}$ and $n_t$ is the index of the activated antenna. Accordingly, the number of bits transmitted in each symbol duration for SM-CIM can be computed as
\begin{equation} \label{eq-SM-CIM-bits}
   b_t = b_1 + b_2 + b_3 = \log_2{N_t} + \log_2{M} + \log_2{N_c}.
\end{equation}

\subsubsection{Receiver}

After perfect carrier recovery and sampling, the received baseband signal $\bm{Y}$ can be expressed as
\begin{align} \label{eq-SM-CIM-received-signal}
   \bm{Y} &= \bm{H} \bm{X} + \bm{N}, 
\end{align}
where $\bm{N}$ is an additive white Gaussian noise (AWGN) matrix, with each entry subject to $\mathcal{CN}(0, N_0)$. Without loss of generality, we assume that the receiver has perfect channel state information (CSI). In principle, we can use the optimal maximum likelihood (ML) detection to jointly detect the index of the activated antenna, the constellation symbols, and the sequence index, i.e.,
\begin{equation} \label{eq-SM-CIM-ML-detection}
   (\hat{n}_c, \hat{n}_t, \hat{x}) = \mathop{\arg\min}\limits_{n_c, n_t, x} \ \left\Vert \bm{Y} - \bm{H} \bm{X}  \right\Vert_F^2.
\end{equation}
The estimated bits $\hat{b}_t$ can be recovered from the estimates $(\hat{n}_c, \hat{n}_t, \hat{x})$. Although ML detection can achieve optimal performance, it requires an exhaustive search over all $N_{c} M N_{t}$ possible transmission matrices, which is computationally intensive. 

To reduce complexity, we can employ the following suboptimal, LC detection method, as illustrated in the right panel of Fig.~\ref{fig-GMAIM-STBC-SM-transceiver}. First, the received baseband signal $\bm{Y}$ performs cross-correlation operation with all the sequences in $\bm{Z}$ to obtain the detection metrics, i.e.,
\begin{align} \label{eq-SM-CIM-detection-metrics}
   \bm{d}_{n_c} = \left\{ \begin{array}{rl}
   \bm{h}_{n_t} E_s x + \bm{N} \bm{z}_{n_c}^{\ast} , & \text{if} \ \hat{n}_c = n_c; \\
   \bm{N} \bm{z}_{n_c}^{\ast}, & \text{otherwise},
   \end{array} \right.
\end{align}
where $\bm{h}_{n_t} \in \mathbb{C}^{N_r \times 1}$ is the $n_t^{\text{th}}$ column of the channel matrix $\bm{H}$, and $E_s \triangleq \bm{z}_{n_c}^H \bm{z}_{n_c}$ denotes as the energy of each symbol. Then, the index of the selected spreading sequence can be estimated as
\begin{align} \label{eq-SM-CIM-detection-codeindex}
   \hat{n}_c = \mathop{\arg\max}\limits_{i \in \{1, 2, \cdots, N_c\}} \ \left\Vert \bm{d}_{i} \right\Vert_2^2.
\end{align}
Next, we use the ML detection to jointly detect the index of the activated antenna and the constellation symbols, i.e.,
\begin{align} \label{eq-SM-CIM-ML-SM}
   (\hat{n}_t, \hat{x}) = \mathop{\arg\min}\limits_{n_t, x} \ \left\Vert \bm{d}_{\hat{n}_c} - \bm{h}_{n_t} x  \right\Vert_2^2.
\end{align}
Finally, the estimated bits $\hat{b}_t$ can be recovered from the estimates $(\hat{n}_c, \hat{n}_t,\hat{x})$. 

\subsection{STBC-SM-CIM Scheme} \label{Section-II-B}
The primary difference in the transceiver design of the STBC-SM-CIM scheme lies in the incorporation of STBC-SM codewords, compared to the SM-CIM scheme. Specifically, the $b_1$ bits are mapped to select the active antenna pair according to the antenna pair combinations summarized in Table~\ref{tab-parameter-STBC-SM}. 

At the transmitter, the constellation symbols corresponding to two consecutive symbol intervals are spread by the same spreading sequence. The transmitted signal matrix is given by 
\begin{equation} \label{eq-STBC-SM-CIM-transmitted-signal-scheme1}
   \bm{X} = e^{j\varphi} \begin{pmatrix}
      \vdots & \vdots \\
      \bm{0}^T & \bm{0}^T \\
      \bm{z}_{n_c}^T x_1 & -\bm{z}_{n_c}^T x_2^{\ast}\\
      \bm{z}_{n_c}^T x_2 & \bm{z}_{n_c}^T x_1^{\ast}\\
      \bm{0}^T & \bm{0}^T \\
      \vdots & \vdots 
   \end{pmatrix},
\end{equation}
where $\bm{X} \in \mathbb{C}^{N_t \times 2L}$, and $x_1$ and $x_2$ are the $M$-ary modulated constellation symbols. The total number of transmitted bits per symbol interval for the STBC-SM-CIM scheme is 
\begin{equation} \label{eq-STBC-SM-CIM-bits-scheme1}
   b_t = b_1 + b_2 + b_3 = \frac{1}{2} \log_2{N_z} + \log_2{M} + \frac{1}{2} \log_2{N_c}.
\end{equation}

At the receiver, the optimal ML detector searches for all possible transmission matrices to jointly detect the constellation symbol, antenna pair index, and spreading sequences index:
\begin{equation} \label{eq-STBC-SM-CIM-ML-detection-scheme1}
   (\hat{n}_c, \hat{n}_z, \hat{x}_1, \hat{x}_2) = \mathop{\arg\min}\limits_{n_c, n_z, x_1, x_2} \ \left\Vert \bm{Y} - \sqrt{\frac{1}{2}} \bm{H} \bm{X}  \right\Vert_F^2.
\end{equation}
However, this exhaustive search requires $N_{c} N_{z} M^2$ metric evaluations, which is computationally prohibitive. To reduce the complexity, we use the following suboptimal detection method. At the receiver, the received baseband signals corresponding to the two symbol intervals are expressed as
\begin{subequations}
\begin{align} \label{eq-STBC-SM-CIM-received-signal-scheme1}
   \bm{Y}_1 &= \sqrt{\frac{1}{2}} \bm{H} \bm{X}_1 + \bm{N}_1, \\
   \bm{Y}_2 &= \sqrt{\frac{1}{2}} \bm{H} \bm{X}_2 + \bm{N}_2, \\
   \bm{Y} &= [\bm{Y}_1 \ \bm{Y}_2],
\end{align}
\end{subequations}
where $\bm{Y}_1, \bm{Y}_2 \in \mathbb{C}^{N_r \times L}$ are the received signals in the first and second symbol intervals, respectively; $\bm{N}_1, \bm{N}_2$ are the corresponding AWGN matrices; and $\bm{X} = [\bm{X}_1 \ \bm{X}_2]$ represents the transmitted signal matrix, where 
\begin{align} \label{eq-STBC-SM-CIM-codeword-scheme1}
   \bm{X}_1 = e^{j\varphi} \begin{pmatrix}
      \vdots \\
      \bm{0}^T \\
      \bm{z}_{n_c}^T x_1 \\
      \bm{z}_{n_c}^T x_2 \\
      \bm{0}^T \\
      \vdots 
   \end{pmatrix},
   \bm{X}_2 = e^{j\varphi} 
   \begin{pmatrix}
      \vdots \\
      \bm{0}^T \\
      -\bm{z}_{n_c}^T x_2^{\ast} \\
      \bm{z}_{n_c}^T x_1^{\ast} \\
      \bm{0}^T \\
      \vdots 
   \end{pmatrix}.
\end{align}
To perform sequence index detection, the receiver first applies cross-correlation between $\bm{Y}_1$ and $\bm{Y}_2$ with each candidate spreading sequence in the codebook $\bm{Z}$, yielding the detection metrics:
\begin{align} \label{eq-STBC-SM-CIM-detection-metrics-scheme1}
   \bm{d}_{n_c,\tau} &= \left\{ \begin{array}{rl}
   \bm{h}_{n_z,1} E_s x_1 + \bm{h}_{n_z,2} E_s x_2 + \bm{N}_{\tau} \bm{z}_{n_c}^{\ast}, & \text{if} \ \hat{n}_c = n_c; \\
   \bm{N}_{\tau} \bm{z}_{n_c}^{\ast}, & \text{otherwise},
   \end{array} \right. 
\end{align}
where $\tau \in \{1,2\}$ denotes the index, and $\bm{h}_{n_z,1}, \bm{h}_{n_z,2} \in \mathbb{C}^{N_r \times 1}$ denote the channel coefficient vectors corresponding to the selected transmit antenna pair with $n_z \in \{1, 2, \cdots, N_z\}$. Based on the obtained detection metrics, the sequence index of the selected spreading sequence can then be estimated as
\begin{align} \label{eq-STBC-SM-CIM-detection-codeindex-scheme1}
   \hat{n}_c = \mathop{\arg\max}\limits_{i \in \{1, 2, \cdots, N_c\}} \ \left(\sum_{\tau=1}^{2} \left\Vert \bm{d}_{i,\tau} \right\Vert_2^2\right).
\end{align}
Next, ML detection is employed to estimate the active antenna-pair index jointly and the transmitted constellation symbols, given the previously detected sequence index $\hat{n}_c$. Following the detection structure in \cite{STBC-SM}, the estimates of $x_1$ and $x_2$ for each antenna pair candidate $p$ are obtained as
\begin{align} \label{eq-STBC-SM-CIM-ML-STBC-SM-scheme1}
   \hat{x}_{p,\tau} &= \mathop{\arg\min}\limits_{x_{\tau}} \ \left\Vert \bm{d} - \sqrt{\frac{1}{2}} \tilde{\bm{h}}_{p,\tau} x_{\tau} \right\Vert_2^2, 
\end{align}
where $\bm{d} = \operatorname{vec}(\bm{D}^T)$, and $\bm{D} = [\bm{d}_{\hat{n}_c,1} \ \bm{d}_{\hat{n}_c,2}^{\ast}]$ is the combined detection matrix for the estimated spreading sequence index $\hat{n}_c$. The equivalent channel matrix for the $p^{\text{th}}$ antenna pair is denoted as $\bm{\mathcal{H}}_p = [\tilde{\bm{h}}_{p,1} \ \tilde{\bm{h}}_{p,2}]$. For instance, when $N_t = 4$, the equivalent channel matrices can be written as
\begin{align} \label{eq-STBC-SM-equivalent-channel-matrix}
   \bm{\mathcal{H}}_1 &=
   \scalebox{0.8}{$\begin{pmatrix}
      h_{1,1} & h_{1,2} \\
      h_{1,2}^{\ast} & -h_{1,1}^{\ast} \\
      h_{2,1} & h_{2,2} \\
      h_{2,2}^{\ast} & -h_{2,1}^{\ast} \\
      \vdots & \vdots \\
      h_{N_r,1} & h_{N_r,2} \\
      h_{N_r,2}^{\ast} & -h_{N_r,1}^{\ast}
   \end{pmatrix}
   $},
   \bm{\mathcal{H}}_2 =
   \scalebox{0.8}{$
   \begin{pmatrix}
      h_{1,3} & h_{1,4} \\
      h_{1,4}^{\ast} & -h_{1,3}^{\ast} \\
      h_{2,3} & h_{2,4} \\
      h_{2,4}^{\ast} & -h_{2,3}^{\ast} \\
      \vdots & \vdots \\
      h_{N_r,3} & h_{N_r,4} \\
      h_{N_r,4}^{\ast} & -h_{N_r,3}^{\ast}
   \end{pmatrix}
   $}, \nonumber \\
   \bm{\mathcal{H}}_3 &=
   \scalebox{0.8}{$
   \begin{pmatrix}
      h_{1,2}\theta & h_{1,3}\theta \\
      h_{1,3}^{\ast} \theta^{\ast} & -h_{1,2}^{\ast} \theta^{\ast} \\
      h_{2,2}\theta & h_{2,3}\theta \\
      h_{2,3}^{\ast} \theta^{\ast} & -h_{2,2}^{\ast} \theta^{\ast} \\
      \vdots & \vdots \\
      h_{N_r,2}\theta & h_{N_r,3}\theta \\
      h_{N_r,3}^{\ast} \theta^{\ast} & -h_{N_r,2}^{\ast} \theta^{\ast}
   \end{pmatrix}
   $}, 
   \bm{\mathcal{H}}_4 =
   \scalebox{0.8}{$
   \begin{pmatrix}
      h_{1,4}\theta & h_{1,1}\theta \\
      h_{1,1}^{\ast} \theta^{\ast} & -h_{1,4}^{\ast} \theta^{\ast} \\
      h_{2,4}\theta & h_{2,1}\theta \\
      h_{2,1}^{\ast} \theta^{\ast} & -h_{2,4}^{\ast} \theta^{\ast} \\
      \vdots & \vdots \\
      h_{N_r,4}\theta & h_{N_r,1}\theta \\
      h_{N_r,1}^{\ast} \theta^{\ast} & -h_{N_r,4}^{\ast} \theta^{\ast}
   \end{pmatrix},
   $}
\end{align}
where $h_{i,j}$ denotes the channel coefficient from the $j^{\text{th}}$ transmit antenna to the $i^{\text{th}}$ receive antenna, and $\theta = e^{j\varphi}$ is the corresponding rotation factor. The corresponding minimum ML metrics for $x_1$ and $x_2$ are calculated as
\begin{subequations}
\begin{align} \label{eq-STBC-SM-CIM-ML-metrics-scheme1}
   m_{p,\tau} &= \mathop{\min}\limits_{x_{\tau}} \ \left\Vert \bm{d} - \sqrt{\frac{1}{2}} \tilde{\bm{h}}_{p,\tau} x_{\tau} \right\Vert_2^2, \\
   \hat{n}_z &= \mathop{\arg\min}\limits_{p} \left(\sum_{\tau=1}^{2} m_{p,\tau}\right), 
\end{align}
\end{subequations}
where $\hat{n}_z$ denotes the index corresponding to the selected antenna pair. Finally, the jointly estimated transmitted symbols are obtained as $(\hat{x}_1, \hat{x}_2) = (\hat{x}_{\hat{n}_z,1}, \hat{x}_{\hat{n}_z,2})$. The entire information bit sequence $\hat{b}_t$ can then be recovered from the estimates $(\hat{n}_c, \hat{n}_z, \hat{x}_1, \hat{x}_2)$. 

\subsection{ESTBC-SM-CIM Scheme} \label{Section-II-C}
The ESTBC-SM-CIM scheme extends both the STBC-SM-CIM and FSCSS-IM schemes \cite{FSCSS-IM}. The key distinction is that ESTBC-SM-CIM uses a pair of spreading sequences to spread the STBC-SM codeword. This sequence pair is selected by choosing two orthogonal or quasi-orthogonal sequences from the codebook $\bm{Z}$. The number of possible spreading sequence pairs is given by $N_s = 2^{\left\lfloor \log_2{\binom{N_c}{2}} \right\rfloor}$. 

At the transmitter, the constellation symbols over two consecutive symbol intervals are spread using the same pair of spreading sequences. Consequently, the transmitted signal matrix can be expressed as
\begin{equation} \label{eq-ESTBC-SM-CIM-transmitted-signal-scheme1}
   \bm{X} = e^{j\varphi} \begin{pmatrix}
      \vdots & \vdots \\
      \bm{0}^T & \bm{0}^T \\
      \bm{z}_{n_{c_1}}^T x_{1}^1 + \bm{z}_{n_{c_2}}^T x_{1}^2 & -\bm{z}_{n_{c_1}}^T x_{2}^1 - \bm{z}_{n_{c_2}}^T x_{2}^2\\
      \bm{z}_{n_{c_1}}^T x_{2}^1 + \bm{z}_{n_{c_2}}^T x_{2}^2 & \bm{z}_{n_{c_1}}^T x_{1}^1 + \bm{z}_{n_{c_2}}^T x_{1}^2\\
      \bm{0}^T & \bm{0}^T \\
      \vdots & \vdots
   \end{pmatrix}, 
\end{equation}
where $n_{c_1}$ and $n_{c_2}$ denote the indices of the selected spreading sequence pair, and $(x_1^{\tau}, x_2^{\tau})$ is the $\tau^{\text{th}}$ transmitted symbol pair within a symbol interval. The total number of bits transmitted per symbol interval is given by 
\begin{equation} \label{eq-ESTBC-SM-CIM-bits-scheme1}
   b_t = b_1 + b_2 + b_3 = \frac{1}{2} \log_2{N_z} + 2 \log_2{M} + \frac{1}{2} \log_2{N_s}.
\end{equation}

At the receiver, the received baseband signals over the two consecutive symbol intervals are expressed as 
\begin{align} \label{eq-ESTBC-SM-CIM-received-signal-scheme1}
   \bm{Y}_{\tau} &= \sqrt{\frac{1}{4}} \bm{H} \bm{X}_{\tau} + \bm{N}_{\tau}. 
\end{align}
Similarly, the joint ML estimation can be performed, i.e.,
\begin{align} \label{eq-ESTBC-SM-CIM-ML-detection-scheme1}
   &(\hat{n}_s, \hat{n}_z, \hat{x}_{\hat{n}_z,1}^1, \hat{x}_{\hat{n}_z,2}^1, \hat{x}_{\hat{n}_z,1}^2, \hat{x}_{\hat{n}_z,2}^2) \nonumber \\ 
    &= \mathop{\arg\min} \ \left\Vert \bm{Y} - \sqrt{\frac{1}{4}} \bm{H} \bm{X}  \right\Vert_F^2.
\end{align}
However, the complexity of joint ML estimation is computationally prohibitive. So we use the suboptimal approach. Subsequently, the received signal $\bm{Y}_{\tau}$ is cross-correlated with all sequences in the codebook $\bm{Z}$ to compute the detection metrics:
\begin{align} \label{eq-ESTBC-SM-CIM-detection-metrics-scheme1}
   \bm{D}_{\tau} &= \bm{Y}_{\tau} \bm{Z}^{\ast}. 
\end{align}
Let $\mathcal{I} = \{1, 2, \cdots, N_c\}$ be the index set of all candidate CIM sequences. The index $\hat{n}_{c_1}$ of the first selected sequence is estimated as
\begin{align} \label{eq-ESTBC-SM-CIM-detection-codeindex-scheme1-1}
   \hat{n}_{c_1} = \mathop{\arg\max}\limits_{i \in \mathcal{I}} \ \left(\sum_{\tau=1}^{2} \left\Vert \bm{d}_{i,\tau} \right\Vert_2^2\right),
\end{align}
where $\bm{d}_{i,\tau}$ denotes the $i^{\text{th}}$ column of the detection matrix $\bm{D}_{\tau}$. Similarly, $\hat{n}_{c_2}$ can be estimated as
\begin{align} \label{eq-ESTBC-SM-CIM-detection-codeindex-scheme1-2}
   \hat{n}_{c_2} = \mathop{\arg\max}\limits_{i \in \mathcal{I} \setminus \{\hat{n}_{c_1}\}} \ \left(\sum_{\tau=1}^{2} \left\Vert \bm{d}_{i,\tau} \right\Vert_2^2 \right),
\end{align}
where $\mathcal{I} \setminus \{\hat{n}_{c_1}\}$ denotes the set of indices excluding $\hat{n}_{c_1}$. The estimated indices $\hat{n}_{c_1}$ and $\hat{n}_{c_2}$ are then sorted in ascending order to determine the final estimated sequence pair, with its corresponding index denoted by $\hat{n}_{s} $. Subsequently, the ML detection is employed to detect the index of the active antenna pair jointly and the transmitted constellation symbols, based on $\hat{n}_{c_1}$ and $\hat{n}_{c_2}$:
\begin{subequations}
\begin{align} \label{eq-ESTBC-SM-CIM-ML-ESTBC-SM-scheme1}
   \hat{x}_{p,1}^{\tau} &= \mathop{\arg\min}\limits_{x_{1}^{\tau}} \ \left\Vert \bm{d}_{\tau} - \sqrt{\frac{1}{4}} \tilde{\bm{h}}_{p,1}^{\tau} x_{1}^{\tau}  \right\Vert_2^2, \\
   \hat{x}_{p,2}^{\tau} &= \mathop{\arg\min}\limits_{x_{2}^{\tau}} \ \left\Vert \bm{d}_{\tau} - \sqrt{\frac{1}{4}} \tilde{\bm{h}}_{p,2}^{\tau} x_{2}^{\tau}  \right\Vert_2^2, 
\end{align}
\end{subequations}
where $\bm{d}_{\tau} = \operatorname{vec}(\tilde{\bm{D}}_{\tau}^T)$ with $\tilde{\bm{D}}_{\tau} = [\bm{d}_{\hat{n}_{c_{\tau}},1} \ \bm{d}^{\ast}_{\hat{n}_{c_{\tau}},2}]$. Here, $\bm{d}_{\hat{n}_{c_{\tau}},1}$ and $\bm{d}_{\hat{n}_{c_{\tau}},2}$ are the $\hat{n}_{c_{\tau}}^{\text{th}}$ column of $\bm{D}_{1}$ and $\bm{D}_{2}$, respectively. Then, the associated minimum ML metrics $m_{p,1}^{\tau}$ and $m_{p,2}^{\tau}$ for $x_{1}^{\tau}$ and $x_{2}^{\tau}$, respectively, can be expressed as
\begin{subequations}
\begin{align} \label{eq-ESTBC-SM-CIM-ML-metrics-scheme1}
   m_{p,1}^{\tau} &= \mathop{\min}\limits_{x_{1}^{\tau}} \ \left\Vert \bm{d}_{\tau} - \sqrt{\frac{1}{4}} \tilde{\bm{h}}_{p,1}^{\tau} x_{1}^{\tau}  \right\Vert_2^2, \\
   m_{p,2}^{\tau} &= \mathop{\min}\limits_{x_{2}^{\tau}} \ \left\Vert \bm{d}_{\tau} - \sqrt{\frac{1}{4}} \tilde{\bm{h}}_{p,2}^{\tau} x_{2}^{\tau}  \right\Vert_2^2.
\end{align}
\end{subequations}
Afterward, the index of the selected codeword can be determined by
\begin{align} \label{eq-ESTBC-SM-CIM-ML-metrics-scheme1-index}
   \hat{n}_z = \mathop{\arg\min}\limits_{p} \ \left(\sum_{\tau=1}^{2} \left(m_{p,1}^{\tau} + m_{p,2}^{\tau}\right)\right).
\end{align}
Finally, the estimated bits $\hat{b}_t$ can be recovered from the estimates $(\hat{n}_s, \hat{n}_z, \hat{x}_{\hat{n}_z,1}^1, \hat{x}_{\hat{n}_z,2}^1, \hat{x}_{\hat{n}_z, 1}^2, \hat{x}_{\hat{n}_z, 2}^2)$.

\section{Performance Analysis} \label{Section-III}
In this section, we first analyze the ABEP performance of the proposed SM-CIM, STBC-SM-CIM, and ESTBC-SM-CIM schemes. Additionally, their data rate is compared with existing schemes, including the traditional SM and STBC-SM, PSK-LoRa \cite{PSK-LoRa}, FSCSS-IM \cite{FSCSS-IM}, MIMO-LoRa \cite{MIMO-LoRa-data-rate}, and 1-Bit-STBC-LoRa \cite{1-Bit-STBC-LoRa}. Finally, the computational complexity of the proposed schemes is evaluated.

\subsection{Average Bit Error Probability (ABEP)} \label{Section-III-A}
\subsubsection{\underline{SM-CIM Scheme}}
For the SM-CIM scheme, the receiver first determines the spreading-sequence index through despreading, followed by antenna index and symbol detection. Accordingly, the overall ABEP is decomposed into contributions from these stages, each weighted by its corresponding bit allocation. Let $P_b$ denote the total ABEP, where $P_1$ represents the ABEP associated with the SM ML detection stage, and $P_2$ denotes the ABEP introduced by the despreading operation. Accordingly, the total ABEP $P_b$ can be expressed as
\begin{align} \label{eq-SM-CIM-ABEP-Pb}
   P_b = \frac{b_1 + b_2}{b_t} P_1 + \frac{b_3}{b_t} P_2.
\end{align}
First, based on the orthogonality or quasi-orthogonality of the spreading sequences, the error probability $P_2$ associated with the despreading process can be computed as \cite[Eq.~(13.4-47)]{Digital-Communication}: 
\begin{align} \label{eq-SM-CIM-ABEP-Pe1}
   P_e &= 1 - \int_{0}^{\infty} \frac{1}{(1+\overline{\gamma}_c)^{N_r} (N_r-1)!} x^{N_r-1} e^{-\frac{x}{1+\overline{\gamma}_c}} \nonumber \\
       &\quad \times \left(1 - e^{-x} \sum_{k=0}^{N_r-1} \frac{x^k}{k!}\right)^{N_c-1} \mathrm{d}x, 
\end{align}
where $P_e$ denotes the average spreading symbol error probability, and $\overline{\gamma}_c = \overline{\gamma}_s / N_r$ is the average SNR per diversity channel. The average SNR per symbol $\overline{\gamma}_s$ is given by 
\begin{align} \label{eq-SM-CIM-ABEP-gamma_s}
   \overline{\gamma}_s = \frac{E_s}{N_0} \sum_{k=1}^{N_r} \mathbb{E}(h_{k}^2),
\end{align}
where $\mathbb{E}(h_{k}^2)$ is the average channel gain of the $k^{\text{th}}$ receive antenna branch, and $h_k \sim \mathcal{CN}(0,1)$. Accordingly, the ABEP of the despreading process can be approximated as 
\begin{align} \label{eq-SM-CIM-ABEP-P2-final}
   P_2 = \frac{N_c}{2N_c-2} P_e.
\end{align}
Next, we analyze the ABEP $P_1$ associated with the ML detection of the SM scheme. According to \cite[Eq.~(4.13)]{SM-ML}, $P_1$ is upper bounded by
\begin{align} \label{eq-SM-CIM-ABEP-P1}
   P_1 \leq \frac{1}{b_t 2^{b_t}} \sum_{i=1}^{b_t} \sum_{q=1}^{b_t} e_{i,q} \alpha^{N_r} \sum_{p=0}^{N_r-1} \binom{N_r-1+p}{p} (1-\alpha)^{p}, 
\end{align}
where $e_{i,q}$ is the number of bit errors between the transmitted symbol $\bm{X}_{i q}$ and the estimated symbol $\bm{X}_{\hat{i} \hat{q}}$. The term $\alpha$ in \eqref{eq-SM-CIM-ABEP-P1} is given by 
\begin{align} \label{eq-SM-CIM-ABEP-alpha}
   \alpha = \frac{1}{2} \left(1 - \sqrt{\frac{\overline{\gamma}/2}{1+\overline{\gamma}/2}} \right), 
\end{align}
where $\overline{\gamma}$ is defined as 
\begin{align} \label{eq-SM-CIM-ABEP-gamma}
   \overline{\gamma} = \frac{E_s}{2N_0} \times \left\{\begin{array}{rl}
      \Vert x_i - x_q \Vert_2^2, & \text{if } i=q; \\
      \Vert x_i \Vert_2^2 + \Vert x_q \Vert_2^2, & \text{otherwise,} \end{array} \right.
\end{align}
with $x_i$ and $x_q$ denoting the respective constellation symbols. Finally, substituting \eqref{eq-SM-CIM-ABEP-P2-final} and \eqref{eq-SM-CIM-ABEP-P1} into the total ABEP expression \eqref{eq-SM-CIM-ABEP-Pb} yields $P_b$.

\begin{table*}[!t]
   \centering
   \caption{Data Rate Comparisons of Different Schemes (Number of Bits per $T_s$)}
   \label{tab-Rt-schemes}
   \renewcommand{\arraystretch}{1.2}
   \begin{tabular}{|>{\centering\arraybackslash}m{0.3cm}
                   |>{\centering\arraybackslash}m{0.3cm}
                   |>{\centering\arraybackslash}m{0.3cm}
                   |>{\centering\arraybackslash}m{0.6cm}
                   |>{\centering\arraybackslash}m{0.6cm}
                   |>{\centering\arraybackslash}m{0.6cm}
                   |>{\centering\arraybackslash}m{0.6cm}
                   |>{\centering\arraybackslash}m{0.6cm}
                   |>{\centering\arraybackslash}m{0.6cm}
                   |>{\centering\arraybackslash}m{1.3cm}
                   |>{\centering\arraybackslash}m{2.0cm}
                   |>{\centering\arraybackslash}m{2.2cm}|}
   \hline
   \multirow{2}{*}{\textbf{$N_t$}} & 
   \multirow{2}{*}{\textbf{$N_c$}} & 
   \multirow{2}{*}{\textbf{$M$}} & 
   \multirow{2}{*}{\textbf{\cite{SM}}} & 
   \multirow{2}{*}{\textbf{\cite{STBC-SM}}} & 
   \multirow{2}{*}{\textbf{\cite{PSK-LoRa}}} & 
   \multirow{2}{*}{\textbf{\cite{FSCSS-IM}}} &
   \multirow{2}{*}{\textbf{\cite{MIMO-LoRa-data-rate}}} &
   \multirow{2}{*}{\textbf{\cite{1-Bit-STBC-LoRa}}} &
   \multirow{2}{*}{\textbf{SM-CIM}} &
   \multirow{2}{*}{\textbf{STBC-SM-CIM}} &
   \multirow{2}{*}{\textbf{ESTBC-SM-CIM}} \\
   & & & & & & & & & & & \\
   \hline
   $4$ & $4$ & $4$ & $4$ & $3$ & $4$ & $2$ & $8$ & $1$ & $6$ & $4$ & $6$ \\
   $4$ & $8$ & $4$ & $4$ & $3$ & $5$ & $4$ & $12$ & $1$ & $7$ & $4.5$ & $7$ \\
   $8$ & $16$ & $8$ & $6$ & $5$ & $7$ & $6$ & $32$ & $1$ & $10$ & $7$ & $11$ \\
   \hline
   \end{tabular}
\end{table*}

\subsubsection{\underline{STBC-SM-CIM Scheme}} \label{Section-III-B}
For the STBC-SM-CIM scheme, transmit diversity introduced by STBC enhances the reliability of the despreading stage. The ABEP derivation follows a structure similar to that of the SM-CIM case, with STBC-aware detection incorporated into the analysis. Similar to \eqref{eq-SM-CIM-ABEP-Pe1}, the ABEP of the despreading process can be computed as 
\begin{align} \label{eq-STBC-SM-CIM-ABEP-P2}
   P_2 &= 1 - \int_{0}^{\infty} \frac{1}{(1+\overline{\gamma}_c)^{2 N_r} (2 N_r-1)!} x^{2 N_r-1} e^{-\frac{x}{1+\overline{\gamma}_c}} \nonumber \\
       &\quad \times \left(1 - e^{-x} \sum_{k=0}^{2 N_r-1} \frac{x^k}{k!}\right)^{N_c-1} \, \mathrm{d}x, 
\end{align}
where the term $2 N_r$ accounts for the diversity gain from using an STBC across multiple receive antennas. The ABEP $P_1$ of the ML detection of the STBC-SM scheme can be upper bounded by \cite{STBC-SM}: 
\begin{align} \label{eq-STBC-SM-CIM-ABEP-P1}
   P_1 \leq \frac{1}{b_t 2^{b_t}} \sum_{i=1}^{b_t} \sum_{q=1}^{b_t} e_{i,q} \text{Pr}\left(\bm{X}_i \to \bm{X}_q\right),
\end{align}
where $e_{i,q}$ denotes the number of bit errors between $\bm{X}_i$ and $\bm{X}_q$, and $\text{Pr}\left(\bm{X}_i \to \bm{X}_q\right)$ is the pairwise error probability (PEP) of incorrectly detecting $\bm{X}_q$ when $\bm{X}_i$ is transmitted. Assuming Rayleigh fading channels, the conditional PEP of the STBC-SM system is given by \cite{GSTBC-SM}
\begin{align} \label{eq-STBC-SM-CIM-ABEP-PEP}
   \text{Pr}\left(\bm{X}_i \to \bm{X}_q | \bm{H} \right) = Q \left( \sqrt{\frac{E_s}{2 N_0} \Vert \bm{H}(\bm{X}_i-\bm{X}_q)\Vert_F^2} \right).
\end{align}
For the convenience of analysis, we use the alternative MGF expression of the Gaussian function \cite{258319}
\begin{align} \label{Q-function-MGF}
    Q(x) = \frac{1}{\pi} \int_{0}^{\frac{\pi}{2}} \exp\left(-\frac{x^2}{2 \sin^2 \theta}\right) \, \mathrm{d}\theta.
\end{align}
In this expression, $\theta$ serves as a dummy variable of integration. However, in certain contexts, it can also be interpreted as a parameter characterizing the correlation between channels or antenna elements, as in MIMO systems. By averaging over the channel statistics using the MGF of a Gaussian distribution, the unconditional PEP becomes 
\begin{align} \label{eq-STBC-SM-CIM-ABEP-PEP-unconditional}
   \text{Pr}\left(\bm{X}_i \to \bm{X}_q \right) &= \mathbb{E}_{\bm{H}} \left( Q \left( \sqrt{\frac{E_s}{2 N_0} \Vert \bm{H}(\bm{X}_i-\bm{X}_q)\Vert_F^2} \right) \right) \nonumber \\
   &= \frac{1}{\pi} \int_{0}^{\frac{\pi}{2}} \int_{-\infty}^{\infty} \exp\left(-\frac{\psi}{2 \sin^2 \theta}\right) f_{\psi}(\psi) \, \mathrm{d}\psi \, \mathrm{d}\theta \nonumber \\
   &= \frac{1}{\pi} \int_{0}^{\frac{\pi}{2}} M_{\psi} \left(-\frac{1}{2 \sin^2 \theta}\right) \, \mathrm{d}\theta,
\end{align}
where $M_{\psi}(\cdot)$ is the MGF of the random variable $\psi = \frac{E_s}{2 N_0} \Vert \bm{H}(\bm{X}_i - \bm{X}_q)\Vert_F^2$. Let the Gram matrix of $\bm{X}_i - \bm{X}_q$ be expressed via eigenvalue decomposition as $\bm{D}_{i,q} = (\bm{X}_i - \bm{X}_q)(\bm{X}_i - \bm{X}_q)^H = \bm{U} \bm{\Lambda}_{i,q} \bm{U}^H$, where $\bm{\Lambda}_{i,q} = \text{diag}(\lambda_{i,q,1}, \lambda_{i,q,2}, \cdots, \lambda_{i,q, N_t})$ contains the non-zero eigenvalues and $r_{i,q}$ is the rank of $\bm{D}_{i,q}$. Defining $\overline{\bm{H}} = \bm{H} \bm{U}$, $\psi$ can be rewritten as 
\begin{align} \label{eq-STBC-SM-CIM-ABEP-PEP-unconditional-psi}
   \psi &= \frac{E_s}{2 N_0} \text{Tr} \left(\bm{H} \bm{D}_{i,q} \bm{H}^H\right) = \frac{E_s}{2 N_0} \text{Tr} \left(\overline{\bm{H}} \bm{\Lambda}_{i,q} \overline{\bm{H}}^H\right) \nonumber \\
   &= \frac{E_s}{2 N_0} \sum_{n=1}^{N_r} \sum_{m=1}^{r_{i,q}} \lambda_{i,q,m} |\overline{h}_{n,m}|^2,
\end{align}
where $\overline{h}_{n,m}$ is the $(n,m)^{\text{th}}$ entry of $\overline{\bm{H}}$. Let $\psi' = \sum_{n=1}^{N_r} \sum_{m=1}^{r_{i,q}} \lambda_{i,q,m} |\overline{h}_{n,m}|^2$ and $\psi_n = \lambda_{i,q,m} |\overline{h}_{n,m}|^2$.  The MGF of $\psi'$ can be computed using the properties of the Chi-squared and exponential distributions \cite[Eq.~(9.12)]{Digital-Communication-fading}:
\begin{align} \label{eq-STBC-SM-CIM-ABEP-PEP-unconditional-psi-MGF}
   M_{\psi'}(x) &= \mathbb{E}_{\psi'} \left(e^{-x \psi'}\right) = \prod_{n=1}^{N_r} \prod_{m=1}^{r_{i,q}} M_{\psi_n}(x) \nonumber \\ 
   &= \prod_{m=1}^{r_{i,q}} (1-\lambda_{i,q,m} x) ^{-N_r}.
\end{align}
Therefore, the MGF of $\psi$ is given by 
\begin{align} \label{eq-STBC-SM-CIM-ABEP-PEP-unconditional-psi-MGF-final}
   M_{\psi}(x) = \prod_{m=1}^{r_{i,q}} M_{\psi'}\left(\frac{E_s}{2 N_0} x\right) = \prod_{m=1}^{r_{i,q}} \left(1-\frac{E_s}{2 N_0} \lambda_{i,q,m} x\right)^{-N_r} \hspace{-0.5em} .
\end{align}
Finally, substituting \eqref{eq-STBC-SM-CIM-ABEP-PEP-unconditional-psi-MGF-final} into \eqref{eq-STBC-SM-CIM-ABEP-PEP-unconditional} yields the closed-form expression of the unconditional PEP:
\begin{align} \label{eq-STBC-SM-CIM-ABEP-PEP-unconditional-final}
\hspace{-0.5em}   \text{Pr}\left(\bm{X}_i \to \bm{X}_q \right) = \frac{1}{\pi} \int_{0}^{\frac{\pi}{2}} \prod_{m=1}^{r_{i,q}} \left(1+\frac{E_s \lambda_{i,q,m}}{4 N_0 \sin^2 \theta}\right)^{-N_r} \hspace{-0.5em} \mathrm{d}\theta .
\end{align}
Moreover, since $(\bm{X}_i - \bm{X}_q) \in \mathbb{C}^{N_t \times 2}$, its rank is at most two. Consequently, both $(\bm{X}_i - \bm{X}_q)(\bm{X}_i - \bm{X}_q)^H \in \mathbb{C}^{N_t \times N_t}$ and $(\bm{X}_i - \bm{X}_q)^H (\bm{X}_i - \bm{X}_q) \in \mathbb{C}^{2 \times 2}$ are Hermitian positive semidefinite matrices. By the singular value decomposition, they share the same nonzero eigenvalues, which are at most two in number. Therefore, \eqref{eq-STBC-SM-CIM-ABEP-PEP-unconditional-final} can be simplified to:
\begin{align} \label{eq-STBC-SM-CIM-ABEP-PEP-unconditional-final-2}
   & \text{Pr}\left(\bm{X}_i \to \bm{X}_q \right) \nonumber \\
   &= \frac{1}{\pi} \int_{0}^{\frac{\pi}{2}} \left(1+\frac{E_s \lambda_{i,q,1}}{4 N_0 \sin^2 \theta}\right)^{-N_r} \left(1+\frac{E_s \lambda_{i,q,2}}{4 N_0 \sin^2 \theta}\right)^{-N_r} \hspace{-0.5em} \mathrm{d}\theta . 
\end{align}
Finally, substituting \eqref{eq-STBC-SM-CIM-ABEP-PEP-unconditional-final} (or its simplified form \eqref{eq-STBC-SM-CIM-ABEP-PEP-unconditional-final-2}) into \eqref{eq-STBC-SM-CIM-ABEP-P1}, and combining it with \eqref{eq-STBC-SM-CIM-ABEP-P2}, the overall ABEP of the STBC-SM-CIM scheme can be obtained via \eqref{eq-SM-CIM-ABEP-Pb}.

\subsubsection{\underline{ESTBC-SM-CIM Scheme}} \label{Section-III-C}
For the ESTBC-SM-CIM scheme, the ABEP is derived under ML detection using a PEP-based framework, where a pair of spreading sequences forms the transmitted matrix. The derivation explicitly traces the sequence from signal construction to the ML decision rule and the corresponding error events. The ABEP of the ESTBC-SM-CIM scheme with ML detection is upper bound by \cite{Digital-Communication-fading}:
\begin{align} \label{eq-ESTBC-SM-CIM-ABEP-Pb}
    P_b \leq \frac{1}{b_t 2^{b_t}} \sum_{i=1}^{b_t} \sum_{q=1}^{b_t} e_{i,q} \text{Pr}\left(\bm{X}_i \to \bm{X}_q\right), 
\end{align}
where $b_t$ is defined in \eqref{eq-ESTBC-SM-CIM-bits-scheme1}, and $\bm{X}$ is the transmitted signal matrix defined in \eqref{eq-ESTBC-SM-CIM-transmitted-signal-scheme1}. The rest of the derivation follows the same approach as that of the STBC-SM-CIM scheme. To further
simplify the calculation of the ABEP upper bound, the value of $Q(x)$ can be approximately computed as \cite{GSTBC-SM}
\begin{align} \label{eq-Qx-approximation}
   Q(x) \approx \frac{1}{12} e^{-\frac{1}{2}x^2} + \frac{1}{4} e^{-\frac{2}{3}x^2},
\end{align}
for large $x$. Then, refer to \eqref{Q-function-MGF}-\eqref{eq-STBC-SM-CIM-ABEP-PEP-unconditional} and \eqref{eq-Qx-approximation}, the corresponding PEP can be upper-bounded by:
\begin{align} \label{eq-ESTBC-SM-CIM-ABEP-PEP-scheme1}
   &\text{Pr}\left(\bm{X}_i \to \bm{X}_q \right) \nonumber \\
   &\approx \int_{0}^{\infty} \left(\frac{1}{12} e^{-\frac{E_s \psi}{4 N_0}} + \frac{1}{4} e^{-\frac{E_s \psi}{3 N_0}}\right) f_{\psi}(\psi) \, \mathrm{d}\psi \nonumber \\
   &= \frac{1}{12} M_{\psi}\left(-\frac{E_s}{4 N_0}\right) + \frac{1}{4} M_{\psi}\left(-\frac{E_s}{3 N_0}\right) \nonumber \\
   &= \frac{1}{12} \prod_{m=1}^{r_{i,q}} \left(1+\frac{E_s \lambda_{i,q,m}}{4 N_0}\right)^{-N_r} \nonumber \\
   &\quad {}+ \frac{1}{4} \prod_{m=1}^{r_{i,q}} \left(1+\frac{E_s \lambda_{i,q,m}}{3 N_0}\right)^{-N_r}.
\end{align} 
It should be emphasized that the ABEP expression for ESTBC-SM-CIM is derived under ML detection and serves as an analytical benchmark. The practical receiver considered for implementation is the proposed LC detector, whose performance relative to ML is evaluated in Section~\ref{Sec-IV-C}.

\subsection{Data Rate and Energy Efficiency}

\begin{table*}[!t]
   \centering
   \caption{Energy Saving Comparisons of Different Schemes}
   \label{tab-EE-schemes}
   \renewcommand{\arraystretch}{1.2}
   \begin{tabular}{|>{\centering\arraybackslash}m{0.3cm}
                   |>{\centering\arraybackslash}m{0.3cm}
                   |>{\centering\arraybackslash}m{0.3cm}
                   |>{\centering\arraybackslash}m{1.0cm}
                   |>{\centering\arraybackslash}m{1.0cm}
                   |>{\centering\arraybackslash}m{1.0cm}
                   |>{\centering\arraybackslash}m{2.0cm}
                   |>{\centering\arraybackslash}m{1.3cm}
                   |>{\centering\arraybackslash}m{2.0cm}
                   |>{\centering\arraybackslash}m{2.2cm}|}
   \hline
   \multirow{2}{*}{\textbf{$N_t$}} & 
   \multirow{2}{*}{\textbf{$N_c$}} & 
   \multirow{2}{*}{\textbf{$M$}} & 
   \multirow{2}{*}{\textbf{\cite{SM}}} & 
   \multirow{2}{*}{\textbf{\cite{STBC-SM}}} & 
   \multirow{2}{*}{\textbf{\cite{PSK-LoRa}}} & 
   \multirow{2}{*}{\textbf{\cite{FSCSS-IM}, \cite{MIMO-LoRa-data-rate}, \cite{1-Bit-STBC-LoRa}}} &
   \multirow{2}{*}{\textbf{SM-CIM}} &
   \multirow{2}{*}{\textbf{STBC-SM-CIM}} &
   \multirow{2}{*}{\textbf{ESTBC-SM-CIM}} \\
   & & & & & & & & & \\
   \hline
   4 & 4 & 4 & $28.57\%$ & $14.28\%$ & $28.57\%$ & $0\%$ & $57.14\%$ & $28.57\%$ & $28.57\%$ \\
   4 & 8 & 4 & $22.22\%$ & $11.11\%$ & $33.33\%$ & $0\%$ &  $55.55\%$ & $27.77\%$ & $33.33\%$ \\
   8 & 16 & 8 & $21.42\%$ & $14.28\%$ & $28.57\%$ & $0\%$ & $50.00\%$ & $28.57\%$ & $35.71\%$ \\
   \hline
   \end{tabular}
\end{table*}

\subsubsection{\underline{Data Rate}}
The data rates of the proposed SM-CIM, STBC-SM-CIM, and ESTBC-SM-CIM schemes can be derived following \cite{GCIM-SM} and are expressed as
\begin{align} \label{eq-Data-Rate}
   R_t = \frac{1 - P_b}{T_s} \, b_t,
\end{align}
where $T_s$ denotes the symbol duration, $P_b$ represents the ABEP, and $b_t$ is the number of bits conveyed per symbol interval. For analytical clarity, we first consider the ideal case with $P_b = 0$.

Table~\ref{tab-Rt-schemes} summarizes the transmitted bits in a symbol duration of the proposed schemes and compares them against benchmark schemes, including conventional SM and STBC-SM, PSK-LoRa \cite{PSK-LoRa}, FSCSS-IM \cite{FSCSS-IM}, MIMO-LoRa \cite{MIMO-LoRa-data-rate}, and 1-Bit-STBC-LoRa \cite{1-Bit-STBC-LoRa}. For instance, under the same symbol duration $T_s$ and configuration $N_t=4$, $N_c=4$, and $M=4$, the number of transmitted bits per symbol duration is: The SM-CIM scheme transmits $6$ bits, the STBC-SM-CIM scheme transmits $4$ bits, the ESTBC-SM-CIM scheme transmits $6$ bits, whereas the SM, STBC-SM, PSK-LoRa, and FSCSS-IM schemes transmit $4$, $3$, $4$, and $2$ bits, respectively. MIMO-LoRa transmits $8$ bits, and 1-Bit-STBC-LoRa transmits $1$ bit. This comprehensive comparison demonstrates the high data rate capability of the proposed schemes.

It is worth noting that pilot overhead reduces the effective throughput of all schemes, particularly in short-packet LPWAN scenarios. Nevertheless, the proposed schemes retain an advantage in payload bits per data symbol, since the antenna-index and code-index bits are conveyed implicitly without requiring higher-order conventional constellations.

\subsubsection{\underline{Energy Efficiency}}
In the proposed SM-CIM, STBC-SM-CIM, and ESTBC-SM-CIM schemes, only the $b_2$ bits are explicitly transmitted using PSK modulation. In contrast, the $b_1$ and $b_3$ bits are conveyed implicitly via antenna indexing and spreading sequence selection, respectively. As a result, the energy required for modulating $b_1 + b_3$ bits is effectively saved. The energy efficiency improvement, expressed as the percentage of energy saved per transmitted $b_t$ bits, can be calculated by \cite{LPWA, SM}
\begin{align} \label{eq-EE} 
   \eta_\text{EE} = \left(1-\frac{b_2}{b_t}\right) \times 100 \%.
\end{align}

Table~\ref{tab-EE-schemes} compares the energy saving of different schemes by maintaining the same total number of transmitted bits $b_t$, while adjusting the modulation order $M$ accordingly. For example, under the configuration $N_t=4$, $N_c=4$, and $M=4$, the SM-CIM scheme saves $57.14\%$ of the energy, while the STBC-SM-CIM and ESTBC-SM-CIM schemes save $28.57\%$ of the energy. In comparison, the conventional schemes provide the following energy savings: the SM, STBC-SM, and PSK-LoRa schemes save $28.57\%$, $14.28\%$, and $28.57\%$ of energy, respectively. However, the FSCSS-IM, MIMO-LoRa, and 1-Bit-STBC-LoRa schemes do not save any energy. The proposed schemes achieve significant energy savings relative to the benchmark schemes, particularly for higher values of $N_t$ and $N_c$.

\subsection{Computational Complexity Analysis}

\begin{table*}[!t]
    \centering
    \caption{Computational Complexity Comparison of Different Schemes (Number of real-number multiplications per symbol duration)}
    \label{tab-complexity-schemes}
    \renewcommand{\arraystretch}{1.2}
    \begin{tabular}{|l|c|}
        \hline
        \textbf{Scheme} & \textbf{Computational Complexity} \\
        \hline
        SM \cite{SM} & $\mathcal{O}(1+8N_rN_tM)$ \\
        STBC-SM \cite{STBC-SM} & $\mathcal{O}(2+16N_rN_zM)$ \\
        PSK-LoRa \cite{PSK-LoRa} & $\mathcal{O}(4L+4LN_c+4M)$ \\
        FSCSS-IM \cite{FSCSS-IM} & $\mathcal{O}(2L+8LN_s)$ \\
        MIMO-LoRa \cite{MIMO-LoRa-data-rate} & $\mathcal{O}(LN_t+(4LN_rN_t + 4LN_r)Nc^{N_t})$ \\
        1-Bit-STBC-LoRa \cite{1-Bit-STBC-LoRa} & $\mathcal{O}(2L+8LN_rN_t)$ \\
        SM-CIM (ML / LC) & $\mathcal{O}(4L + (4LN_rN_t + 4LN_r)2^{b_t})$ / $\mathcal{O}(4L + 8LN_cN_r + 8N_rN_tM)$ \\
        STBC-SM-CIM (ML / LC) & $\mathcal{O}(8L + (4LN_rN_t + 4LN_r)2^{b_t})$ / $\mathcal{O}(8L + 4LN_cN_r + 4N_cN_r + 16N_rN_zM)$ \\
        ESTBC-SM-CIM (ML / LC) & $\mathcal{O}(16L + (4LN_tN_r + 4LN_r)2^{b_t})$ / $\mathcal{O}(16L + 4LN_cN_r + 4N_cN_r + 32N_rN_zM)$ \\
        \hline
    \end{tabular}
\end{table*}

We evaluate the computational complexity of the proposed schemes in terms of the number of real-number multiplications required per symbol duration. 
\subsubsection{\underline{SM-CIM Scheme}}
At the transmitter, the SM-CIM scheme involves a single spreading operation, with complexity $\mathcal{O}(4L)$. As for the optimal ML detection, the total complexity is computed as $\mathcal{O}(4L + (4 L N_r N_t + 4 L N_r) 2^{b_t})$. Regarding the LC approach, despreading at the receiver requires $\mathcal{O}(8 L N_c N_r)$ real-number multiplications, whereas SM ML detection entails $\mathcal{O}(8 N_r N_t M)$ operations. Therefore, the total complexity of the LC detection is $\mathcal{O}(4L + 8 L N_c N_r + 8 N_r N_t M)$.

\subsubsection{\underline{STBC-SM-CIM Scheme}} 
The transmitter performs two spreading operations, resulting in a complexity of $\mathcal{O}(8L)$. As for the optimal ML detection, the total computational complexity is computed as $\mathcal{O}(8L + (4 L N_r N_t + 4 L N_r) 2^{b_t})$. As for the LC detection, the despreading complexity is $\mathcal{O}(4 L N_c N_r + 2 N_c N_r)$, and the STBC-SM ML detection requires $\mathcal{O}(16 N_r N_z M)$ multiplications, where $N_z$ denotes the number of candidate STBC-SM codewords. Consequently, the total computational complexity reduces to $\mathcal{O}(8 L + 4 L N_c N_r + 4 N_c N_r + 16 N_r N_z M)$.

\subsubsection{\underline{ESTBC-SM-CIM Scheme}}
In the ESTBC-SM-CIM scheme, four spreading operations are performed at the transmitter, with computational complexity $\mathcal{O}(16L)$. The optimal joint ML detection complexity is computed as $\mathcal{O}(16 L + (4 L N_t N_r + 4 L N_r) 2^{b_t})$. As for the LC detection, the despreading process incurs $\mathcal{O}(4 L N_c N_r + 4 N_c N_r)$ operations, and the STBC-SM ML detection stage has a complexity of $\mathcal{O}(32 N_r N_z M)$. Therefore, the overall computational complexity reduces to $\mathcal{O}(16 L + 4 L N_c N_r + 4 N_c N_r + 32 N_r N_z M)$.

Table~\ref{tab-complexity-schemes} presents a comprehensive comparison of the computational complexity of the proposed SM-CIM, STBC-SM-CIM, and ESTBC-SM-CIM schemes against several benchmark schemes, including conventional SM and STBC-SM, PSK-LoRa, FSCSS-IM, MIMO-LoRa, and 1-Bit-STBC-LoRa. The complexities are evaluated for both the transmitter and receiver sides, with a focus on the number of real-number multiplications required per symbol duration. Compared to the traditional SM and STBC-SM schemes, the proposed SM-CIM and STBC-SM-CIM schemes introduce additional spreading and despreading operations. Compared with PSK-LoRa, SM-CIM incorporates SM processing, whereas STBC-SM-CIM integrates SM and STBC processing. In comparison with FSCSS-IM, the ESTBC-SM-CIM scheme further includes an STBC-SM component. Although the proposed schemes entail higher computational complexity than the benchmark counterparts, they offer significant performance improvements in terms of data rate and energy efficiency, reflecting the fundamental trade-off between complexity and performance.

\section{Simulation Results and Discussion} \label{Section-IV}
This section presents and analyzes simulation and analytical results for the proposed SM-CIM, STBC-SM-CIM, and ESTBC-SM-CIM schemes over Rayleigh fading channels. The performance of these schemes is compared against several benchmark techniques, including conventional SM, STBC-SM, PSK-LoRa \cite{PSK-LoRa} (in which the phase shift is treated as the modulation symbol and the FSCSS sequence serves as the spreading code), FSCSS-IM \cite{FSCSS-IM}, MIMO-LoRa \cite{MIMO-LoRa-data-rate}, and 1-Bit-STBC-LoRa \cite{1-Bit-STBC-LoRa}. All simulations are conducted in MATLAB R2024b. For clarity, we use the CPM-SS sequence \cite{CPM-SS-Long} as the spreading code in our simulation experiments, and the chip sampling rate is set to $P=4$, the spreading factor is set to $\text{SF}=6$, so the length of the CPM-SS sequence is $L = P (2^{\text{SF}}-1) = 252$. The longer the spreading length, the more spreading sequences available for selection, the more bits available for sequence index transmission, and the greater the spreading gain; however, it also increases computational complexity.

In the simulation results, the horizontal axis is $E_s/N_0$, representing the SNR per symbol. When interpreting the effective SNR at the receiver, the spreading gain of $10\log_{10}L \approx 24$ dB should be accounted for, yielding SNR values consistent with LPWAN scenarios. The use of $E_s/N_0$ facilitates direct comparison with theoretical BER expressions and is standard in spreading-system analysis.

\begin{figure*}[!t]
    \centering
    \subfigure[SM-CIM]{
		\includegraphics[width=0.3\linewidth]{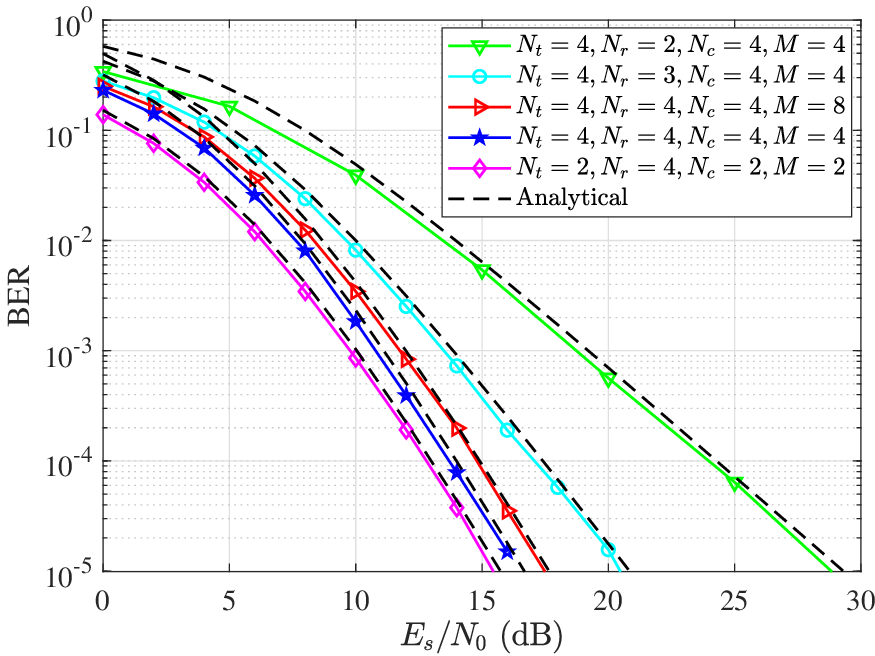} \label{subfig-ber-sm-cess}}
	\subfigure[STBC-SM-CIM]{
		\includegraphics[width=0.3\linewidth]{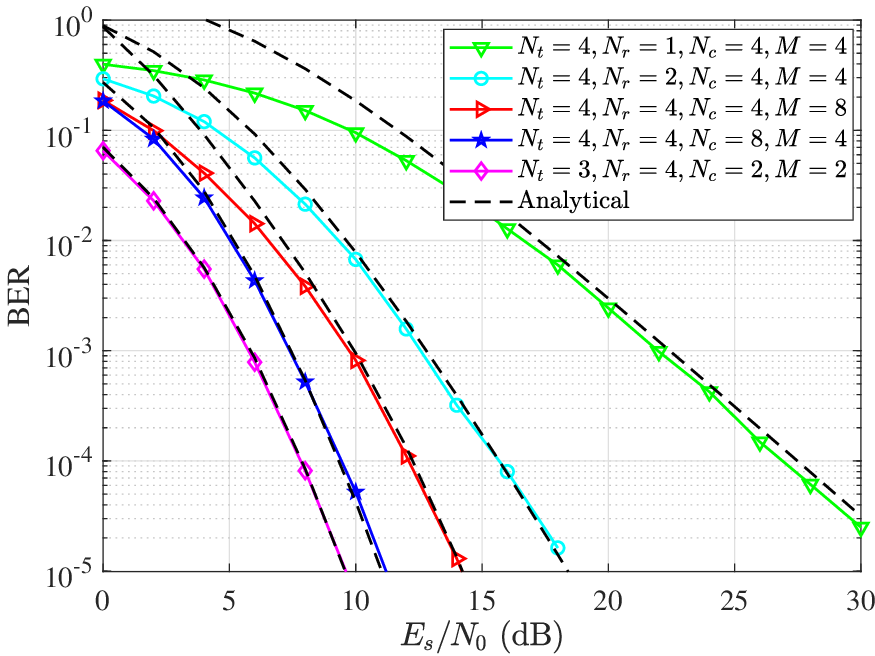} \label{subfig-ber-stbc-sm-cess-1}}
	\subfigure[ESTBC-SM-CIM]{
		\includegraphics[width=0.3\linewidth]{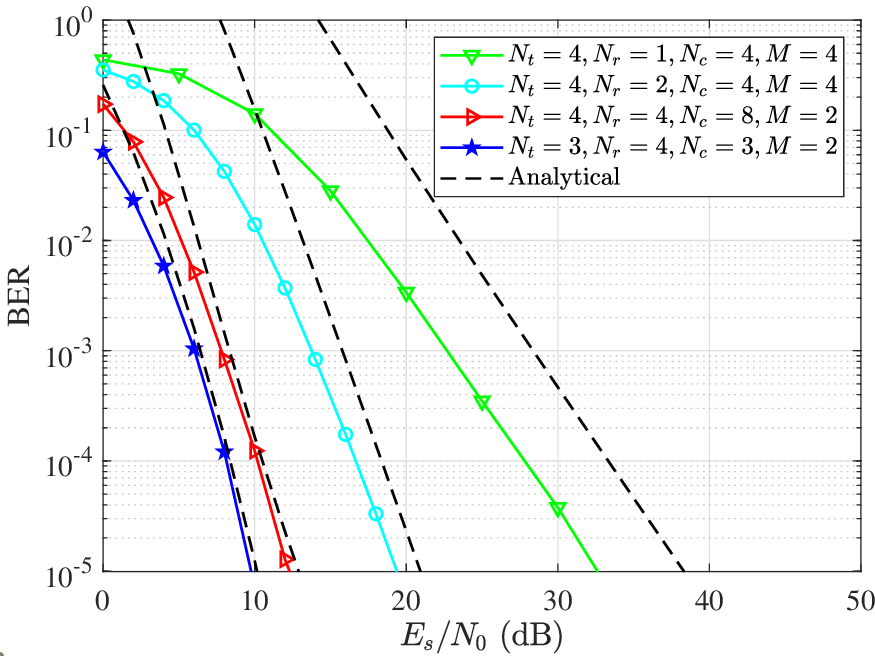} \label{subfig-ber-ESTBC-SM-CIM-1}}
	\caption{BER vs. $E_s/N_0$ of three proposed schemes for different parameters over Rayleigh fading channels.}
\label{fig-ber-proposed-three-schemes}
\end{figure*}

\begin{figure*}[!t]
    \centering
    \subfigure[]{
		\includegraphics[width=0.3\linewidth]{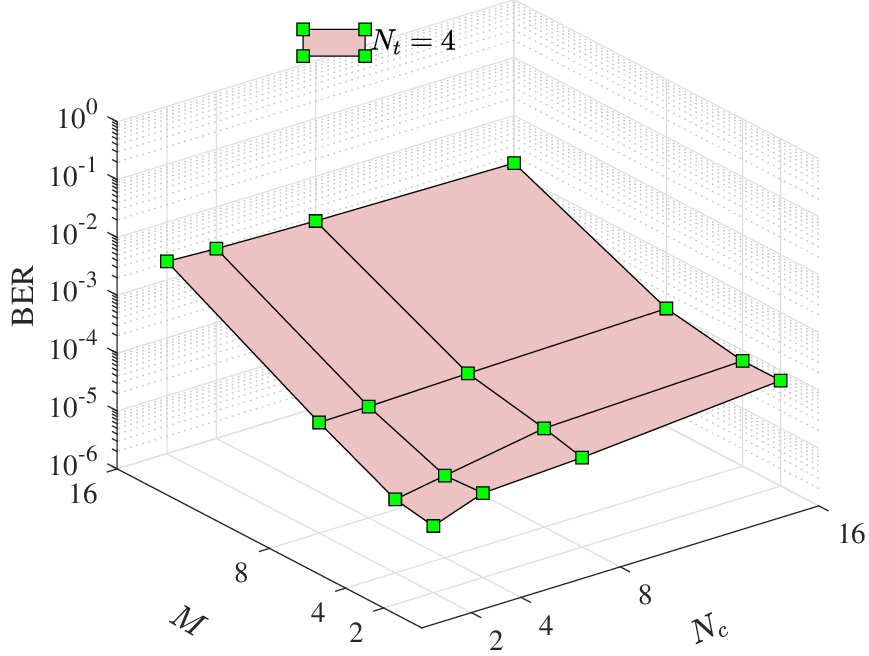} \label{subfig-sm-cess-p-Nt4}}
	\subfigure[]{
		\includegraphics[width=0.3\linewidth]{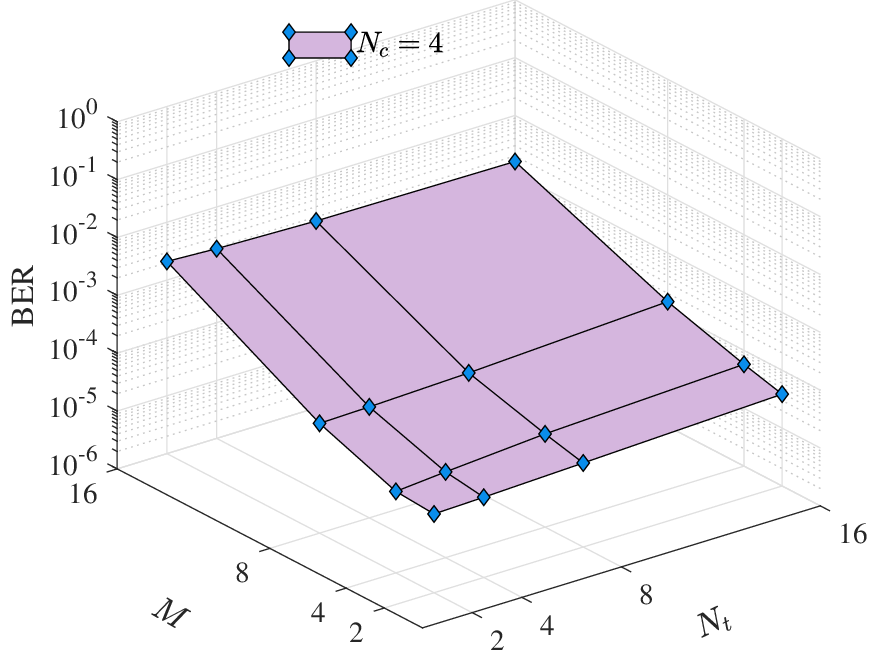} \label{subfig-sm-cess-p-Nc4}}
	\subfigure[]{
		\includegraphics[width=0.3\linewidth]{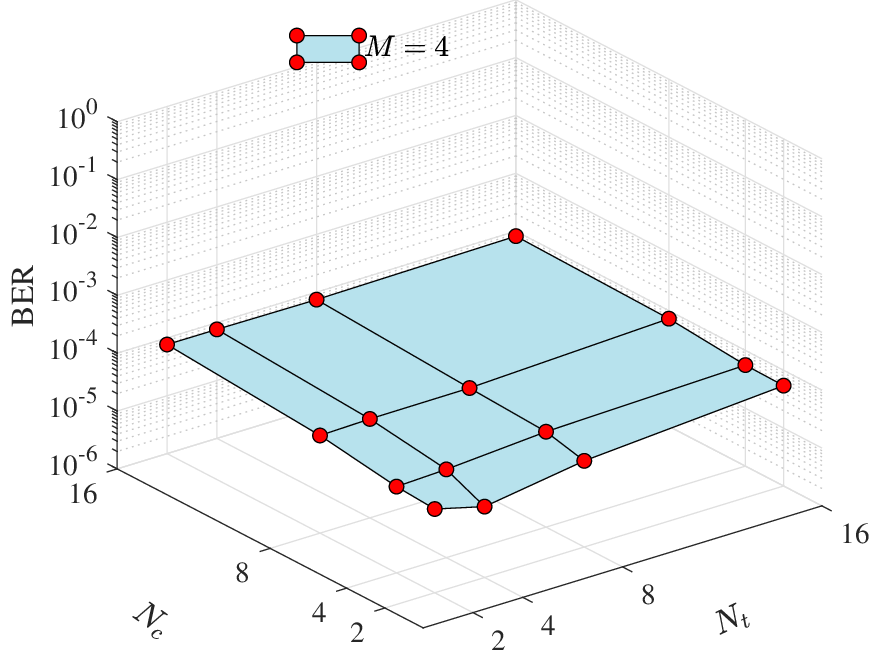} \label{subfig-sm-cess-p-M4}}
	\caption{BER performance of SM-CIM for different parameters with $N_r=4$ when $E_s/N_0 = 15$ \si{dB} over Rayleigh fading channels.}
\label{fig-sm-cess-parameters}
\end{figure*}

\subsection{Analytical ABEP Upper Bound}
Fig.~\ref{fig-ber-proposed-three-schemes} presents the analytical ABEP upper bounds and simulated BER performance of the proposed SM-CIM, STBC-SM-CIM, and ESTBC-SM-CIM schemes under different parameter settings:
\begin{itemize}
    \item SM-CIM Scheme (Fig.~\ref{subfig-ber-sm-cess}): $(N_t, N_r, N_c, M) \in \{(4, 2, 4, 4), (4, 3, 4, 4), (4, 4, 4, 8), (4, 4, 4, 4), (2, 4, 2, 2)\}$, corresponding to $b_t \in \{3, 3, 3.5, 3, 1.5\}$ bits per symbol;
    \item STBC-SM-CIM Scheme (Fig.~\ref{subfig-ber-stbc-sm-cess-1}): $(N_t, N_r, N_c, M) \in \{(4, 1, 4, 4), (4, 2, 4, 4), (4, 4, 4, 8), (4, 4, 8, 4), (4, 4, 2, 2)\}$, corresponding to $b_t \in \{4, 4, 5, 4.5, 2\}$ bits per symbol;
    \item ESTBC-SM-CIM Scheme (Fig.~\ref{subfig-ber-ESTBC-SM-CIM-1}): $(N_t, N_r, N_c, M) \in \{(4, 1, 4, 4), (4, 2, 4, 4), (4, 4, 8, 2), (3, 4, 3, 2)\}$, corresponding to $b_t = \{6, 6, 5, 3\}$ bits per symbol.
\end{itemize}

The analytical curves plotted in Fig.~\ref{fig-ber-proposed-three-schemes} are obtained using the closed-form ABEP expressions derived in Section~\ref{Section-III-A}, i.e., SM-CIM and STBC-SM-CIM corresponding to a LC receiver algorithm, ESTBC-SM-CIM corresponding to an ML receiver algorithm. At the same time, the simulation results are averaged over $10^6$ independent channel realizations. The analytical and simulation results exhibit a close agreement across all schemes and configurations, particularly at high $E_s/N_0$, thereby validating the accuracy of the derived ABEP expressions. Moreover, the BER performance improves with increasing $N_r$ due to higher diversity gain, as reflected in the steeper slopes of the BER curves.

\subsection{BER for Different Parameters}
Fig.~\ref{fig-sm-cess-parameters} depicts the BER performance of the {\it SM-CIM} scheme under different parameter configurations with $N_r = 4$ at $E_s/N_0 = 15$~\si{dB} over Rayleigh fading channels. Specifically, Fig.~\ref{subfig-sm-cess-p-Nt4} shows the impact of varying $M \in \{2,4,8,16\}$ and $N_c \in \{2,4,8,16\}$ with fixed $N_t = 4$. Fig.~\ref{subfig-sm-cess-p-Nc4} illustrates the effect of changing $N_t \in \{2,4,8,16\}$ and $M$ with fixed $N_c=4$, while Fig.~\ref{subfig-sm-cess-p-M4} examines the influence of varying $N_t$ and $N_c$ with fixed $M=4$. The results indicate that increasing $N_t$ or $N_c$ has a relatively minor negative impact on BER, whereas increasing $M$ substantially degrades BER.  

\begin{figure*}[!t]
    \centering
    \subfigure[]{
		\includegraphics[width=0.3\linewidth]{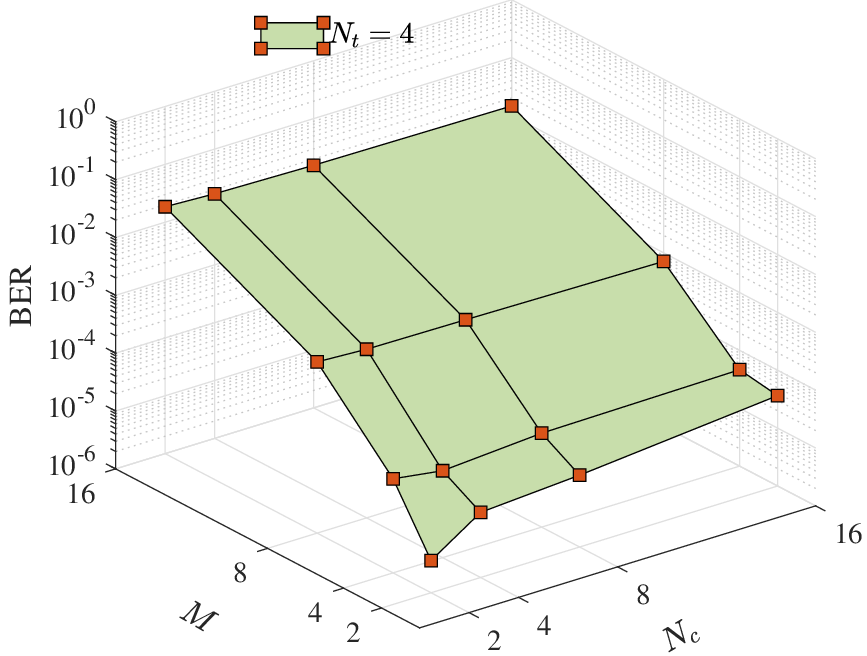} \label{subfig-stbc-sm-cess-1-p-Nt4}}
	\subfigure[]{
		\includegraphics[width=0.3\linewidth]{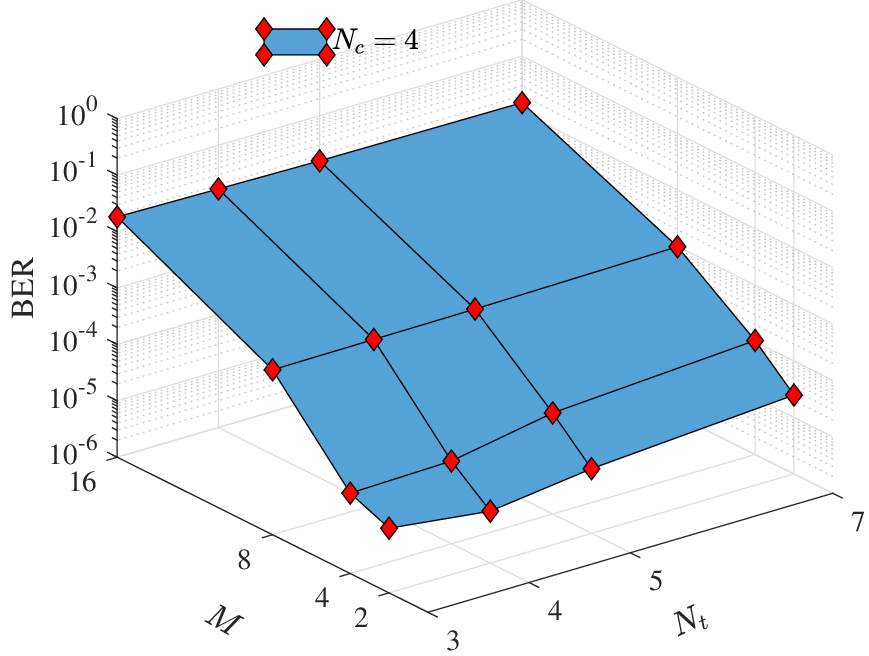} \label{subfig-stbc-sm-cess-1-p-Nc4}}
	\subfigure[]{
		\includegraphics[width=0.3\linewidth]{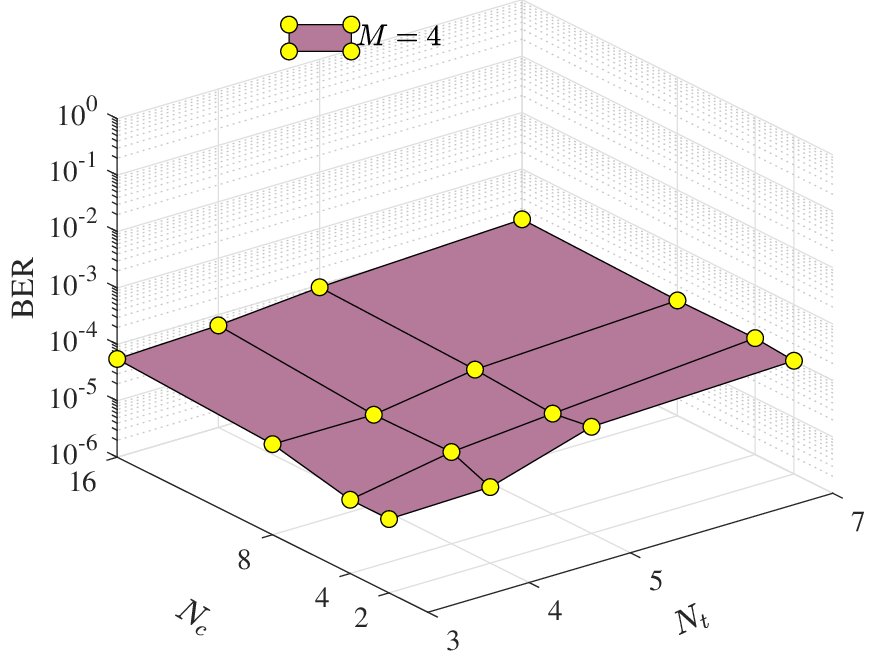} \label{subfig-stbc-sm-cess-1-p-M4}}
	\caption{BER performance of STBC-SM-CIM for different parameters with $N_r=4$ when $E_s/N_0 = 10$ \si{dB} over Rayleigh fading channels.}
\label{fig-stbc-sm-cess-1-parameters}
\end{figure*}

Fig.~\ref{fig-stbc-sm-cess-1-parameters} presents the BER performance of the {\it STBC-SM-CIM} scheme with $N_r = 4$ at $E_s/N_0 = 10$~\si{dB}. 
As shown in Fig.~\ref{subfig-stbc-sm-cess-1-p-Nt4}, varying $M \in \{2,4,8,16\}$ and $N_c \in \{2,4,8,16\}$ with fixed $N_t = 4$ affects the BER performance. Fig.~\ref{subfig-stbc-sm-cess-1-p-Nc4} illustrates the effect of varying $N_t \in \{3,4,5,7\}$ and $M$ with fixed $N_c=4$, while Fig.~\ref{subfig-stbc-sm-cess-1-p-M4} shows the influence of varying $N_t$ and $N_c$ with fixed $M=4$. The results reveal that increasing $N_t$ or $N_c$ slightly degrades BER performance, whereas increasing $M$ significantly degrades it.

\begin{figure*}[!t]
    \centering
    \subfigure[]{
		\includegraphics[width=0.3\linewidth]{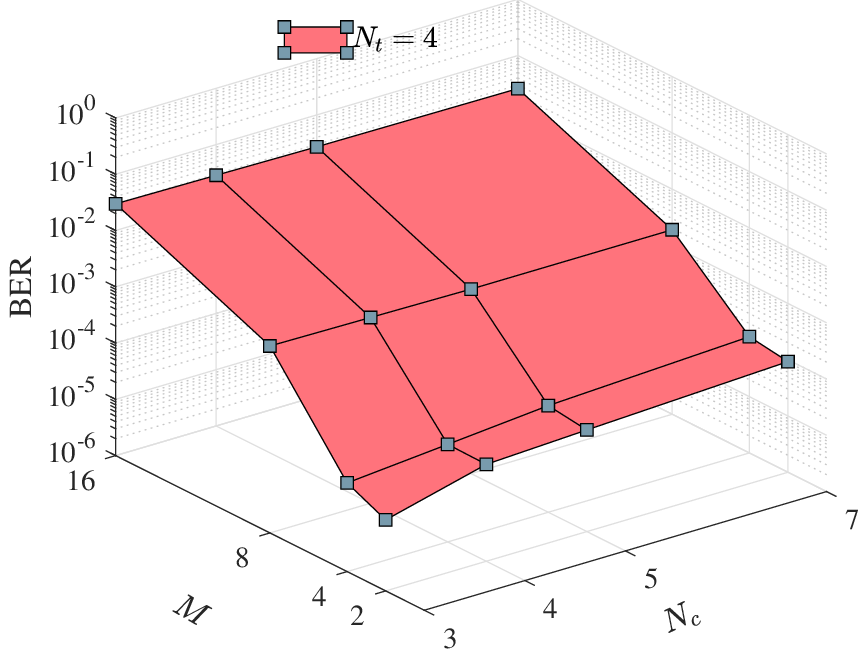} \label{subfig-ESTBC-SM-CIM-1-p-Nt4}}
	\subfigure[]{
		\includegraphics[width=0.3\linewidth]{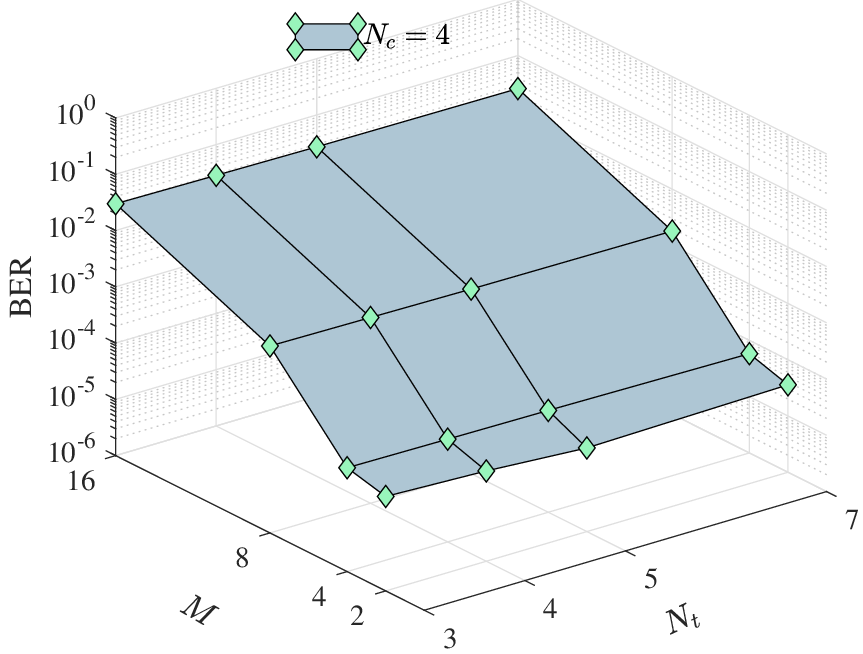} \label{subfig-ESTBC-SM-CIM-1-p-Nc4}}
	\subfigure[]{
		\includegraphics[width=0.3\linewidth]{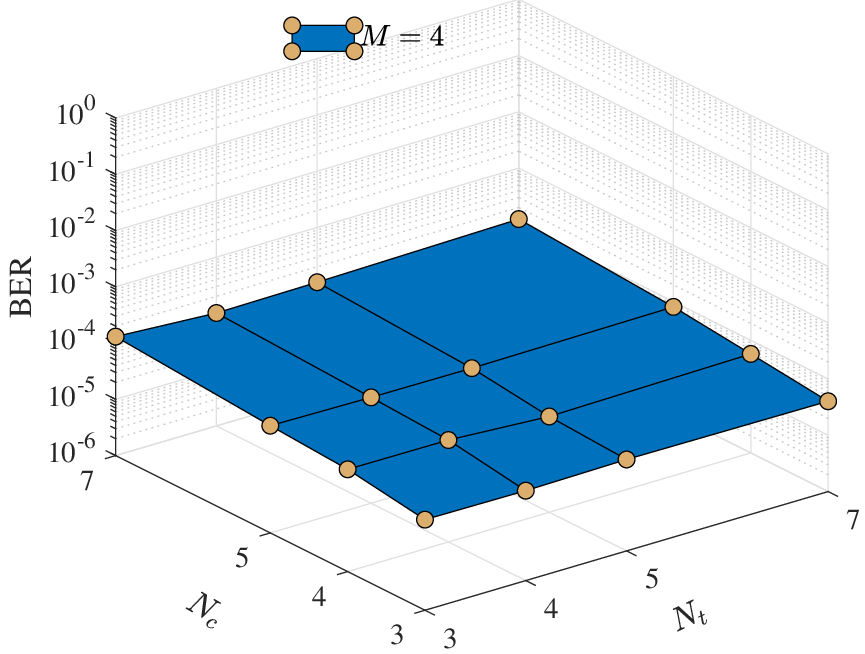} \label{subfig-ESTBC-SM-CIM-1-p-M4}}
	\caption{BER performance of ESTBC-SM-CIM for different parameters with $N_r=4$ when $E_s/N_0 = 12$ \si{dB} over Rayleigh fading channels.}
\label{fig-ESTBC-SM-CIM-1-parameters}
\end{figure*}

Fig.~\ref{fig-ESTBC-SM-CIM-1-parameters} shows the BER performance of the {\it ESTBC-SM-CIM} scheme with $N_r = 4$ at $E_s/N_0 = 12$~\si{dB}. 
In Fig.~\ref{subfig-ESTBC-SM-CIM-1-p-Nt4}, $M \in \{2,4,8,16\}$ and $N_c \in \{3,4,5,7\}$ are varied with fixed $N_t = 4$. Fig.~\ref{subfig-ESTBC-SM-CIM-1-p-Nc4} evaluates the impact of $N_t \in \{3,4,5,7\}$ and $M$ with fixed $N_c=4$, while Fig.~\ref{subfig-ESTBC-SM-CIM-1-p-M4} investigates the influence of varying $N_t$ and $N_c$ with fixed $M=4$. The results demonstrate that increasing $N_t$ or $N_c$ degrades BER performance. Moreover, increasing $M$ from $2$ to $4$ slightly degrades BER performance, whereas further increasing $M$ to $8$ or $16$ causes a significant degradation.  

In summary, the system data rate can be enhanced by adjusting $N_t$, $N_c$, and $M$. However, increasing $N_t$ requires additional transmit antennas, which are often constrained by device dimensions and implementation complexity. Increasing $M$ substantially degrades the BER performance due to the higher modulation sensitivity. By contrast, increasing $N_c$ offers a more favorable trade-off, as it enhances the data rate with relatively limited impact on BER performance.

\subsection{BER Comparison Between ML and LC Detection} \label{Sec-IV-C}
\begin{figure*}[!t]
    \centering
    \subfigure[SM-CIM]{
		\includegraphics[width=0.3\linewidth]{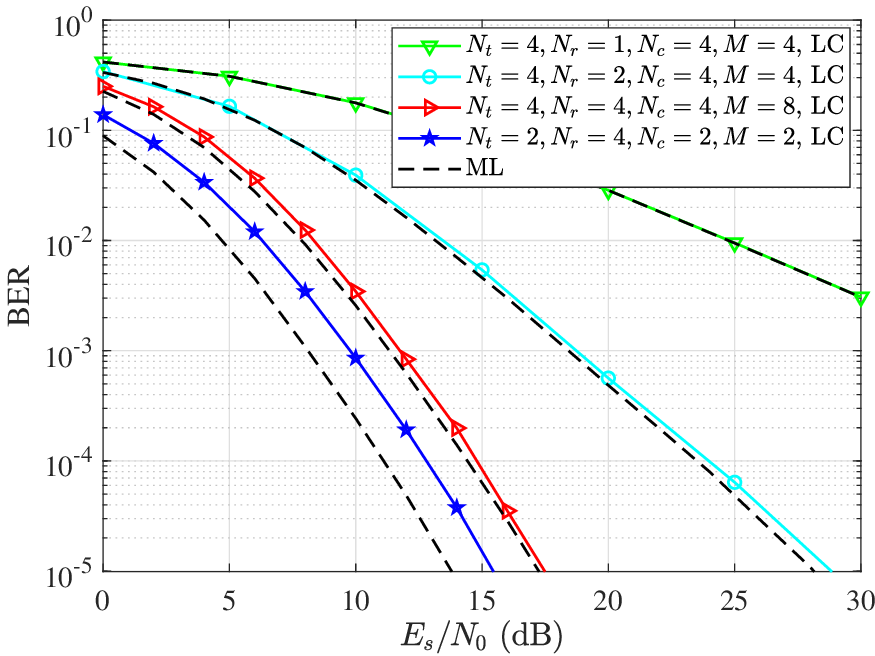} \label{subfig-SM-CIM-ML-LC}
	}
	\subfigure[STBC-SM-CIM]{
		\includegraphics[width=0.3\linewidth]{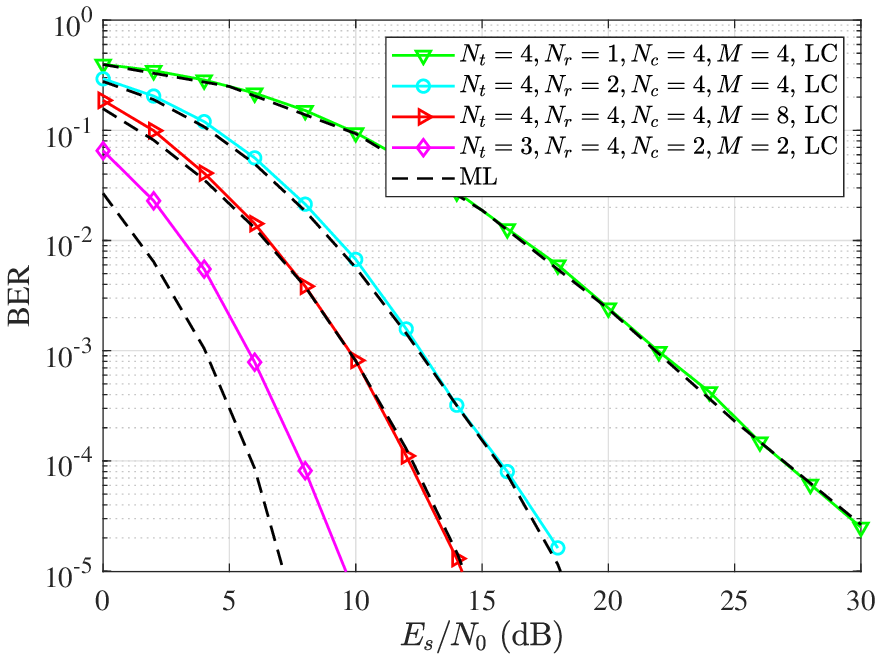} \label{subfig-STBC-SM-CIM-1-ML-LC}
	}
	\subfigure[ESTBC-SM-CIM]{
		\includegraphics[width=0.3\linewidth]{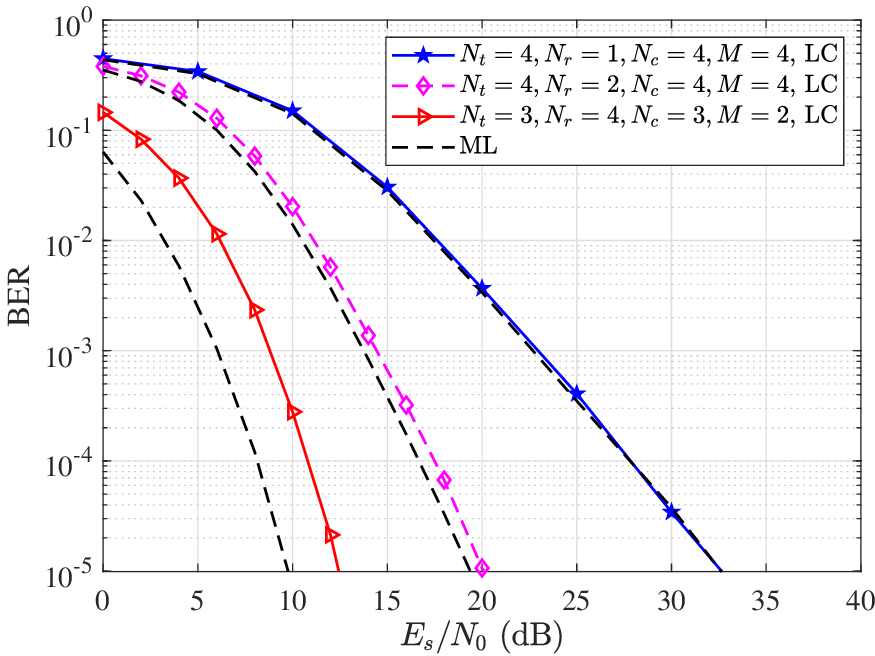} \label{subfig-ESTBC-SM-CIM-1-ML-LC}
	}
	\caption{Comparison of BER performance between ML and LC detection for the proposed SM-CIM, STBC-SM-CIM, and ESTBC-SM-CIM schemes over Rayleigh fading channels.}
   \label{fig-ML-LC-comparison}
\end{figure*}

Fig.~\ref{subfig-SM-CIM-ML-LC} compares the BER performance of ML detection and LC detection for the SM-CIM scheme over Rayleigh fading channels. When the transmission parameters are set to $(N_t, N_r, N_c, M) = (2, 4, 2, 2)$, the LC detection suffers from approximately $1.5$~\si{dB} degradation compared to ML detection. For other transmission-parameter settings, the LC detector's performance approaches that of the ML detector.

Fig.~\ref{subfig-STBC-SM-CIM-1-ML-LC} presents the BER performance comparison between ML and LC detection for the STBC-SM-CIM scheme over Rayleigh fading channels. With transmission parameters $(N_t, N_r, N_c, M) = (3, 4, 3, 2)$, the LC detector experiences about $2$~\si{dB} performance loss relative to ML detection. Under other parameter settings, the two detection schemes exhibit nearly identical performance.

Fig.~\ref{subfig-ESTBC-SM-CIM-1-ML-LC} shows the BER performance comparison between ML and LC detection for the ESTBC-SM-CIM scheme over Rayleigh fading channels. When configured with $(N_t, N_r, N_c, M) = (3, 4, 3, 2)$, the LC detection incurs approximately $2.5$~\si{dB} degradation compared to ML detection. For other transmission configurations, its performance remains close to that of the ML detector.

As shown in Fig.~\ref{fig-ML-LC-comparison}, the LC detector incurs approximately $1.5$--$2.5$ dB loss in the worst-case setting compared to ML detection, while remaining close to ML performance for the other investigated configurations.

\subsection{BER Comparison Between Different Schemes}
We employ the proposed LC detector to evaluate the BER performance of the proposed schemes and compare them with benchmark schemes, including conventional SM, STBC-SM, PSK-LoRa, FSCSS-IM, MIMO-LoRa, and 1-Bit-STBC-LoRa. The performance is assessed in terms of BER versus $E_s/N_0$ under different configurations. 

\subsubsection{\underline{BER Comparison with Basic Schemes}}
\begin{figure}[!t]
    \centering
    \includegraphics[width=0.85\linewidth]{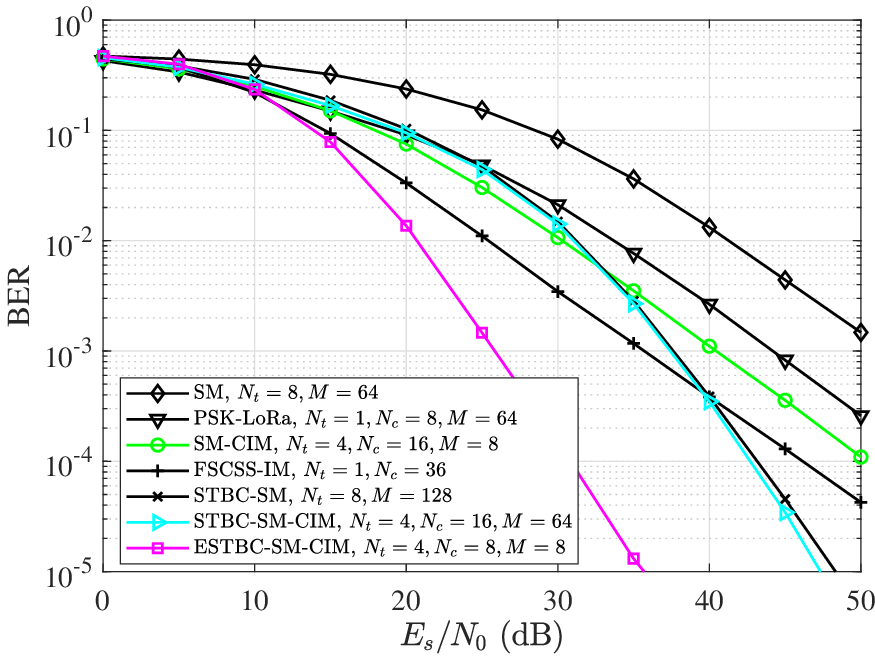}
    \caption{BER performance comparison of the proposed schemes with benchmark schemes over Rayleigh fading channels for $b_t=9$ bits when $N_r=1$.}
    \label{fig-comparison-Nr1}
\end{figure}

Fig.~\ref{fig-comparison-Nr1} compares the BER performance of the proposed schemes with benchmark schemes over Rayleigh fading channels for $b_t=9$ bits and $N_r=1$. The SM scheme is configured with $(N_t, M) = (8, 64)$; the PSK-LoRa scheme with $(N_t, N_c, M) = (1, 8, 64)$; the STBC-SM scheme with $(N_t, M) = (8, 128)$; and the FSCSS-IM scheme with $(N_t, N_c) = (1, 36)$. The proposed SM-CIM scheme is configured with $(N_t, N_c, M) = (4, 16, 8)$; the STBC-SM-CIM with $(N_t, N_c, M) = (4, 16, 64)$; and the ESTBC-SM-CIM with $(N_t, N_c, M) = (4, 8, 8)$.

At $\text{BER} = 10^{-4}$, the BER performance (from worst to best) is observed as: SM, PSK-LoRa, SM-CIM, FSCSS-IM, STBC-SM, STBC-SM-CIM, and ESTBC-SM-CIM. Specifically, the proposed SM-CIM scheme outperforms both SM and PSK-LoRa. The FSCSS-IM scheme surpasses SM-CIM by about $5$~\si{dB}. The STBC-SM-CIM scheme achieves nearly the same performance as STBC-SM. At low $E_s/N_0$, both schemes exhibit performance degradation relative to FSCSS-IM, whereas at high $E_s/N_0$, they outperform FSCSS-IM. Notably, the ESTBC-SM-CIM scheme achieves about $13$~\si{dB} gain over STBC-SM-CIM.

\begin{figure}[!t]
    \centering
    \includegraphics[width=0.85\linewidth]{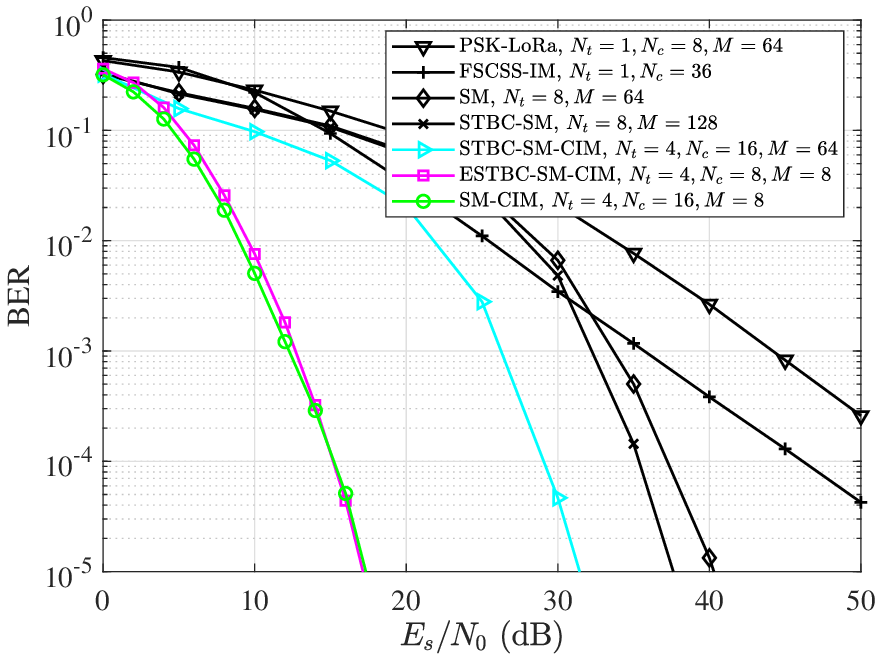}
    \caption{BER performance comparison of the proposed schemes with basic schemes over Rayleigh fading channels for $b_t=9$ bits when $N_r=4$.}
    \label{fig-comparison-Nr4}
\end{figure}

Fig.~\ref{fig-comparison-Nr4} compares BER performance when $N_r=4$. The transmission parameters are identical to those in Fig.~\ref{fig-comparison-Nr1} except for $N_r=4$. At $\text{BER} = 10^{-4}$, the performance ranking (from worst to best) is: PSK-LoRa, FSCSS-IM, SM, STBC-SM, STBC-SM-CIM, ESTBC-SM-CIM, and SM-CIM. Since PSK-LoRa and FSCSS-IM are single-antenna schemes, they lack antenna diversity. Specifically, FSCSS-IM outperforms PSK-LoRa by about $10$~\si{dB}, SM outperforms FSCSS-IM by about $8$~\si{dB}, and STBC-SM outperforms SM by about $2$~\si{dB}. The proposed STBC-SM-CIM achieves approximately $7$~\si{dB} more gain than STBC-SM. Meanwhile, ESTBC-SM-CIM performs comparably to SM-CIM and outperforms STBC-SM-CIM by about $14$~\si{dB}.

\subsubsection{\underline{BER Comparison with State-of-The-Art Schemes}}
We further compare the proposed schemes with state-of-the-art schemes, including MIMO-LoRa \cite{MIMO-LoRa-data-rate} and 1-Bit-STBC-LoRa \cite{1-Bit-STBC-LoRa}, in both perfect-CSI and imperfect-CSI scenarios. The estimated channel coefficient is modeled as $\hat{h}_{n,m} = h_{n,m} + h_{n,m}^{e}$, where $h_{n,m}^{e} \sim \mathcal{CN}(0, \sigma_{\varepsilon}^2)$ represents the estimation error. A larger $\sigma_{\varepsilon}^2$ implies more severe CSI inaccuracy.

\begin{figure}[!t]
    \centering
    \includegraphics[width=0.85\linewidth]{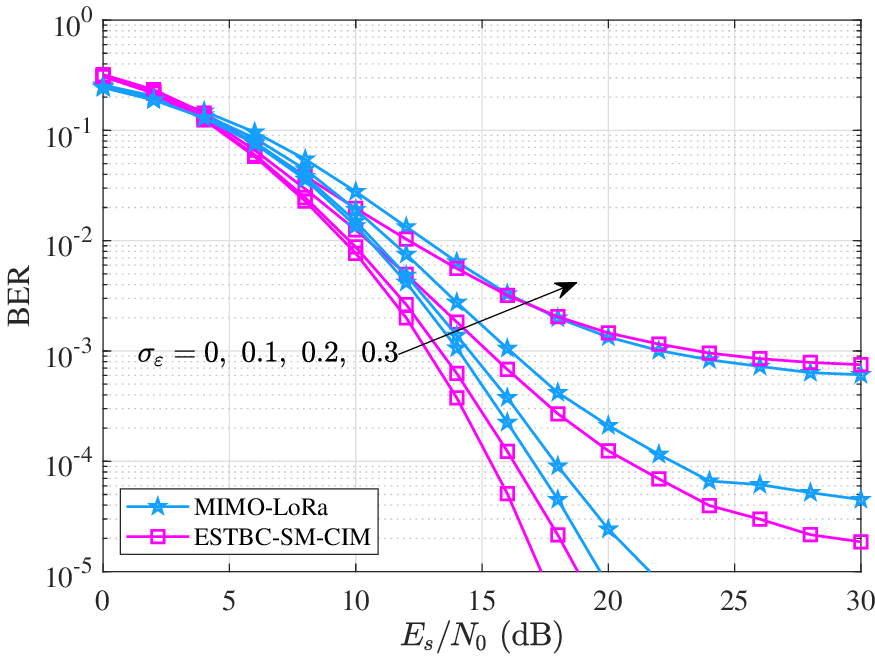}
    \caption{BER performance comparison of the ESTBC-SM-CIM scheme with MIMO-LoRa in imperfect CSI scenarios.}
    \label{fig-comparison-es-MIMO-LoRa}
\end{figure}

Fig.~\ref{fig-comparison-es-MIMO-LoRa} compares the ESTBC-SM-CIM scheme and MIMO-LoRa under imperfect CSI. MIMO-LoRa is configured with 
$(N_t, N_r, N_c) = (4, 4, 4)$, while ESTBC-SM-CIM is configured with $(N_t, N_r, N_c, M) = (4, 4, 4, 8)$ (corresponding to $b_t=8$). At $\text{BER} = 10^{-4}$, ESTBC-SM-CIM achieves about $2$~\si{dB} gain over MIMO-LoRa in the perfect CSI case ($\sigma_{\varepsilon}^2 = 0$). Under imperfect CSI, both schemes degrade as $\sigma_{\varepsilon}^2$ increases, but ESTBC-SM-CIM consistently outperforms MIMO-LoRa.

\begin{figure}[!t]
    \centering
    \includegraphics[width=0.85\linewidth]{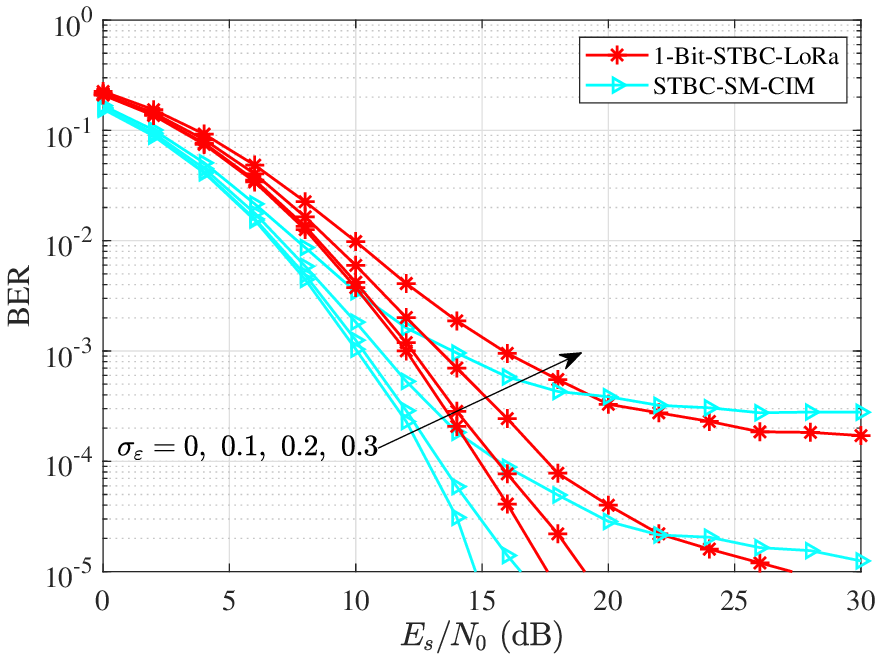}
    \caption{BER performance comparison of the STBC-SM-CIM scheme with 1-Bit-STBC-LoRa in imperfect CSI scenarios.}
    \label{fig-comparison-es-1-bit-STBC-LoRa}
\end{figure}

Fig.~\ref{fig-comparison-es-1-bit-STBC-LoRa} compares STBC-SM-CIM with 1-Bit-STBC-LoRa. The 1-Bit-STBC-LoRa scheme is configured with 
$(N_t, N_r, N_c) = (2, 2, 2)$ (corresponding to $b_t=1$), while STBC-SM-CIM is configured with $(N_t, N_r, N_c, M) = (3, 2, 2, 2)$ (corresponding to $b_t=2$). For fairness, STBC-SM-CIM also activates two transmit antennas and uses the smallest transmission parameters. At $\text{BER} = 10^{-4}$, STBC-SM-CIM achieves about $2$~\si{dB} gain over 1-Bit-STBC-LoRa in the perfect CSI case. Under imperfect CSI, both schemes degrade, but STBC-SM-CIM consistently outperforms 1-Bit-STBC-LoRa. With larger channel estimation errors, the BER performance of 1-Bit-STBC-LoRa at higher $E_s/N_0$ values slightly outperforms that of the proposed STBC-SM-CIM scheme. In most other cases, the proposed scheme achieves better BER performance. Overall, the proposed framework offers a more favorable trade-off between reliability and spectral efficiency under imperfect CSI. Moreover, STBC-SM-CIM transmits $2$ bits per symbol, compared with $1$ bit in 1-Bit-STBC-LoRa, yielding a significant data-rate gain while maintaining a lower BER under challenging conditions. 

Next, we consider correlated MIMO channels, modeled as $\bm{H}_{\text{corr}} = \bm{R}_{r}^{1/2} \bm{H} \bm{R}_{t}^{1/2}$, where $\bm{R}_{r} = [r_{ij}]_{N_r \times N_r}$ and $\bm{R}_{t} = [r_{ij}]_{N_t \times N_t}$ denote the receive and transmit correlation matrices, respectively, and $\bm{H}$ represents the uncorrelated channel matrix \cite{10214216}. The correlation coefficients are modeled as $r_{ij} = r_{ji}^{\ast} = r^{|j-i|}$, where $r \in [0,1]$ denotes the correlation strength. This model captures the spatial correlation arising from antenna proximity, where the correlation between antenna elements decreases with increasing separation. A larger value of $r$ indicates stronger spatial correlation, thereby reducing channel independence, degrading diversity gain, and leading to performance loss. In contrast, when $r$ is small, the system approaches the uncorrelated MIMO case. These effects are consistent with prior studies \cite{STBC-SM, 10214216}.

\begin{figure}[!t]
    \centering
    \includegraphics[width=0.85\linewidth]{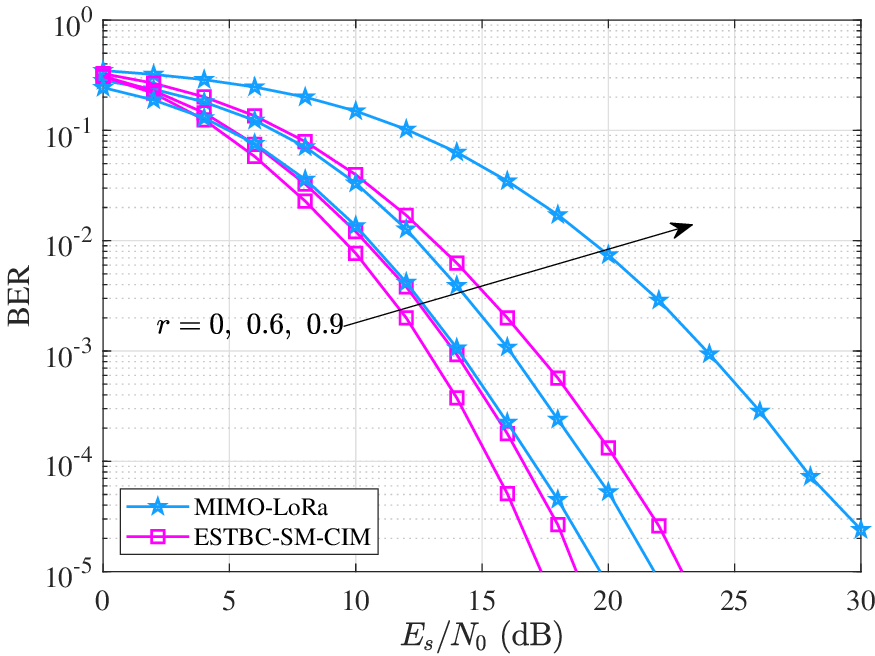}
    \caption{BER performance comparison of the ESTBC-SM-CIM scheme with MIMO-LoRa in correlated channel scenarios.}
    \label{fig-comparison-corr-MIMO-LoRa}
\end{figure}

\begin{figure}[!t]
    \centering
    \includegraphics[width=0.85\linewidth]{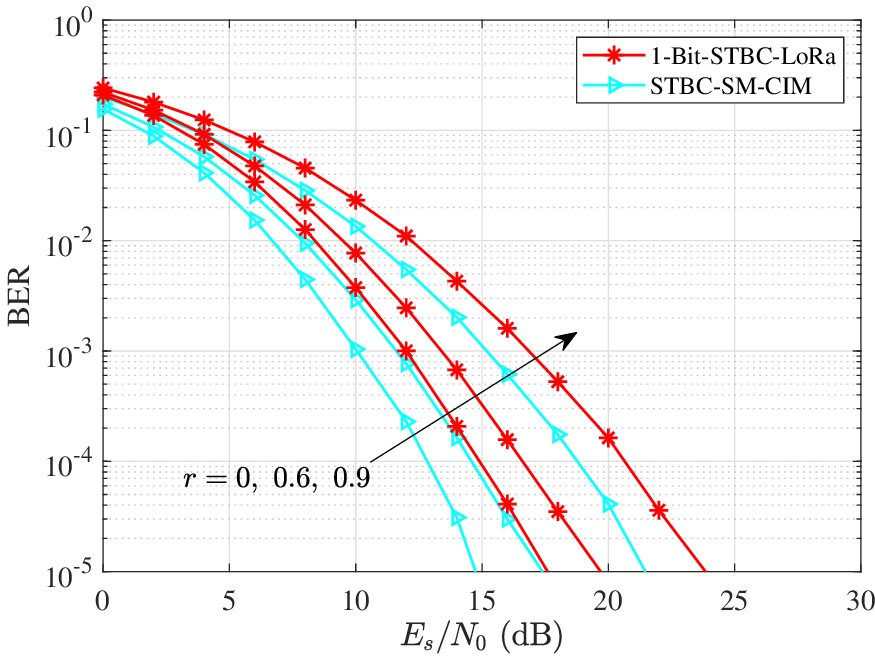}
    \caption{BER performance comparison of the STBC-SM-CIM scheme with 1-Bit-STBC-LoRa in correlated channel scenarios.}
    \label{fig-comparison-corr-1-bit-STBC-LoRa}
\end{figure}

Fig.~\ref{fig-comparison-corr-MIMO-LoRa} shows the BER performance comparison between ESTBC-SM-CIM and MIMO-LoRa under correlated channels. With $(N_t, N_r, N_c) = (4, 4, 4)$ (corresponding to $b_t=8$), ESTBC-SM-CIM consistently outperforms MIMO-LoRa. At $\text{BER} = 10^{-4}$, both schemes degrade with increasing correlation $r$, but ESTBC-SM-CIM provides about $8$~\si{dB} gain over MIMO-LoRa when $r=0.9$, demonstrating its robustness to severe channel correlation. Fig.~\ref{fig-comparison-corr-1-bit-STBC-LoRa} compares STBC-SM-CIM with 1-Bit-STBC-LoRa in correlated channels. Both schemes degrade as the correlation $r$ increases, but STBC-SM-CIM consistently achieves lower BER. 

\subsection{Advantages and Limitations of the Proposed Schemes}
In summary, the proposed schemes consistently outperform benchmark approaches, offering notable improvements in BER, data rate, and energy efficiency, and are robust to imperfect CSI and correlated channels. Furthermore, they allow flexible adjustment of the number of bits per symbol, enabling a tunable trade-off between throughput and power consumption. From a system perspective, the proposed LC detector is essential to the scalability of the code--spatial framework. By decoupling the despreading stage from subsequent antenna and symbol detection, it avoids the exhaustive joint search required by ML detection and provides a scalable alternative, particularly in large-$N_r$ or high-$b_t$ regimes, while maintaining near-ML performance.

It is worth noting that the pilot overhead reduces the absolute throughput of all schemes, particularly in short-packet LPWAN scenarios, which is a significant consideration for our future work. For analytical tractability, the channel is assumed to follow a block-fading model, i.e., it remains constant over one packet (or, equivalently, over the two consecutive symbol intervals comprising each STBC codeword). This assumption is well-suited to low-mobility LPWAN scenarios. In time-selective channels with non-negligible Doppler effect, channel variations within a packet may degrade the performance of despreading and STBC decoding. Extending the proposed framework to Doppler-aware channel tracking and time-varying CSI estimation is left for future work.

\section{Conclusion} \label{Section-V}
This paper develops a unified code--spatial index modulation framework for LPWAN communications, in which SM-CIM, STBC-SM-CIM, and ESTBC-SM-CIM represent different operating points on a common diversity--rate--complexity tradeoff surface. By jointly exploiting antenna selection, spreading-sequence selection, and space-time coding, the framework enhances data rate and energy efficiency while maintaining practical detectability through the proposed LC receiver. The advantages become more pronounced in regimes with larger $N_r$ or higher $b_t$, where the LC detector remains near-ML while exhaustive ML detection becomes increasingly impractical. The framework supports flexible bits-per-symbol adjustment, enabling a tunable throughput-power trade-off, and offers a practical, scalable solution for high-rate, energy-efficient LPWAN communications. Future work will extend to channel estimation and massive IoT scenarios, and investigate performance under high-mobility scenarios.

\bibliographystyle{IEEEtran}
\bibliography{MyRef}

\end{document}